\documentclass[twocolumn,amssymb, nobibnotes, floatfix, aps, prb]{revtex4-2}
\usepackage{hyperref}
\usepackage[utf8]{inputenc}
\usepackage[english]{babel}
\usepackage{amssymb}
\usepackage{amsmath}
\usepackage{graphicx}
\usepackage{color}
\usepackage{physics}
\usepackage{float}
\usepackage{bbold}
\renewcommand{\vec}[1]{\boldsymbol{#1}}
\graphicspath{{Figs/}}
\newcommand{\be}{\begin{equation}}
\newcommand{\ee}{\end{equation}}
\newcommand{\bea}{\begin{eqnarray}}
\newcommand{\eea}{\end{eqnarray}}

\def\nn{\nonumber}
\def\lb{\label}
\def\pref{\eqref}

\begin{document}


\title{Coulomb interaction, phonons, and superconductivity in twisted bilayer graphene.}

\author{Tommaso Cea$^{1,2}$, and Francisco Guinea$^{1,3,4}$}
\affiliation{
$^{1}$ 
Imdea Nanoscience, Faraday 9, 28015 Madrid, Spain.
\\
$^{2}$
Instituto de Ciencia de Materiales de Madrid. Consejo Superior de Investigaciones Cient{\'\i}ficas, Sor Juana In\'es de la Cruz 3. 28049 Madrid. Spain.
\\
$^{3}$
Donostia International Physics Center, Paseo Manuel de Lardiz\'abal 4, 20018 San Sebasti\'an, Spain.
\\
$^{4}$
Ikerbasque Basque Foundation for Science, Bilbao, Spain. 
}

\date{\today}

\begin{abstract}
The polarizability of twisted bilayer graphene, due to the combined effect of electron-hole pairs, plasmons, and acoustic phonons is analyzed. The screened Coulomb interaction allows for the formation of Cooper pairs and superconductivity in a significant range of twist angles and fillings. The tendency toward superconductivity is enhanced by the coupling between longitudinal phonons and electron-hole pairs. Scattering processes involving large momentum transfers, Umklapp processes, play a crucial role in the formation of Cooper pairs. The magnitude of the superconducting gap changes among the different pockets of the Fermi surface.
\end{abstract}


\maketitle
\section{Introduction}
\label{sec:intro}

Twisted bilayer graphene (TBG) shows a complex phase diagram, which combines superconducting and insulating phases\cite{Cao2018,cao_nat18}, and resembles strongly correlated materials previously encountered in condensed matter physics\cite{D77,S84,S11,KKNUZ15}. On the other hand, superconductivity seems more prevalent in TBG\cite{Letal19,Setal20,Aetal20,Setal20b,Cetal21}, while in other strongly correlated materials magnetic phases are dominant.

The pairing interaction responsible for superconductivity in TBG has been intensively studied. 
Among other possible pairing mechanisms, the effect of phonons\cite{Peltonen_prb18,WMM18,CC18,LWB19,WHS19,SAO20,QZM21,choi_choi_cm21} (see also\cite{ATF19}), the proximity of the chemical potential to a van Hove singularity in the density of states (DOS)\cite{IYF18,SB18,GS19,CCC20,LN20}, and excitations of insulating phases\cite{PZVS18,YV19,KZB20} (see also\cite{KBVZ20,KXM20,WS20}), or the role of electronic screening\cite{RJ19,GCML19,LCR20,STSVA20}, have been considered.

In the following, we analyze how the screened Coulomb interaction induces pairing in TBG. The calculation is based on the Kohn-Luttinger formalism\cite{KL_prl65} for the study of anisotropic superconductivity via repulsive interactions. The screening includes electron-hole pairs\cite{Petal19}, plasmons\cite{LL19}, and phonons (note that acoustic phonons overlap with the electron-hole continuum in TBG). Our results show that the repulsive Coulomb interaction, screened by plasmons and electron-hole pairs only, leads to anisotropic superconductivity, although with critical temperatures of order $T_c \sim 10^{-3} - 10^{-2}$K. The inclusion of phonons in the screening function substantially enhances the critical temperature, to $T_c \sim 1 - 10$ K. 

\section{The model}
\label{sec:model}
\subsection{Electronic structure and electron-electron interactions}
The long range Coulomb interaction, projected onto the central bands of TBG, is described by an energy scale in the range of 20-100 meV.
As a result, this interaction modifies significantly the shape and width of the bands of TBG near the first magic angle. The Hartree potential widens the bands, as it shifts the energies at the $K$ and $M$ points of the moiré Brillouin zone (BZ) with respect to those at the $\Gamma$ point\cite{Guinea_pnas18,cea_prb19,RAM19,CB20,GVLML20}. The inclusion of the exchange term in a full Hartree-Fock calculation leads to broken symmetry phases, with valley and/or spin polarization\cite{XM20,CG20,ZJVZ20,LKYV21}, among other possible phases (see also \cite{LZCY18,ZJWZ20,Betal20,LD21,LKYV21}). 

In the following, we consider the role of long range charge fluctuations in the superconducting pairing of twisted bilayer graphene. These fluctuations couple to quasiparticles, and also among themselves, via the long range Coulomb interactions. In addition, long range fluctuations can be induced by longitudinal acoustic phonons (see below, and also\cite{KL_prl65}), so that we include these phonons in the calculation. 
\\
We analyze pairing by charge fluctuations using the leading contributions in perturbation theory, already considered in\cite{KL_prl65}. We describe, however, the charge fluctuations using the RPA response function, as the large density of states and the spin and valley multiplicities imply a significant renormalization of these excitations.
\\
We analyze phases without broken symmetries, where pairing occurs between electrons in different valleys. This analysis can be extended to broken symmetry phases where the two valleys are partially occupied, as expected\cite{Zetal20,CG20} for fillings $\nu = \pm 2$.
\\
We do not consider here pairing due to optical phonons, or to acoustic phonons which do not induce long wavelength charge fluctuations. We discuss later whether these interactions enhance, or suppress, the pairing analyzed here.
\\
The self consistent screening of the long range interactions, mediated by the long wavelength charge excitations, is approximated by the static response function. The neglect of retardation effects, imposed by the complexity of the calculation, implies that pairing is overestimated when the obtained critical temperature is comparable to the frequencies where the response function is expected to have significant structure, such as the phonon frequencies. 


The electronic properties of the model are described by the continuum model for TBG\cite{LPN07,BM11}, using the parameters in\cite{koshino_prx18}, $\hbar v_F = 5.18$ eV \AA , $\{g_{AA} , g_{AB} \} = \{ 0.0797 \, {\rm eV} , 0.0975 \, {\rm eV} \}$ (see the App. \ref{APP_continuum_model} for further details). The difference between $g_{AA}$ and $g_{AB}$, as described in\cite{koshino_prx18}, accounts for the inhomogeneous interlayer distance, which is minimum in the $AB/BA$ regions and maximum in the $AA$ ones, or it can be seen as a model of a more complete 
 treatment of lattice relaxation\cite{Guinea2019}.

\subsection{Electron-phonon interactions}
We consider as well the electron-electron interaction mediated by acoustic phonons. The energy of these phonons, for wavelengths similar to the moiré unit cell, are of the order of 1-4 meV, comparable to the energy of electron-hole interactions. We focus on longitudinal phonons, which couple to electrons via the deformation potential, $D \sim 20$ eV. We neglect the interaction between phonons in the two layers, and consider the even superposition of a phonon in each layer such that the displacements in the two layers have the same sign. We neglect the odd superposition, which induces shifts in the chemical potential of the two layers of opposite sign, so that it does not lead to a net charge accumulation. The exchange of a phonon leads to an effective, momentum and frequency dependent interaction between electrons: 
\begin{align}
{\cal V}_{eff}^{ph} ( \vec{q} , \omega ) &= \frac{D^2 |\vec{q}|^2}
{\rho ( \omega^2 - \omega_{\vec{q}}^2 )},
\end{align}
where $\rho$ is the mass density, $\omega_{\vec{q}} = v_s | \vec{q} |$
is the frequency of a phonon of momentum $\vec{q}$ and $v_s=\sqrt{( \lambda + 2 \mu )/\rho} $ is the velocity of sound, $\lambda$ and $\mu$ being the elastic Lam\'e coefficients. At low frequencies
the electron-phonon coupling, in a single graphene layer, can be described by a dimensionless parameter which describes the phonon induced electron-electron interaction: $\tilde{g} = D^2 / ( \lambda + 2 \mu ) \times |\chi_0| $, where $\chi_0$ is the electronic susceptibility at zero frequency and momentum (see also\cite{LNC21}). 

We use $\lambda + 2 \mu \approx 20$ eV \AA$^{-2}$ and $\chi_0 =- {\cal D} ( \epsilon_F ) \sim 4 \times ( W A_C )^{-1} \approx 2 \times 10^{-2}$ eV$^{-1}$ \AA$^{-2}$, where ${\cal D} ( \epsilon_F )$ is the DOS at the Fermi energy, $W \sim 10$ meV is the electron bandwidth, $A_C \sim 2 \times 10^4$ \AA$^2$ is the area of the unit cell, and the factor 4 stands for the spin and valley degeneracy. Then, the dimensionless electron-phonon interaction is $\tilde{g} \approx 0.4$. For a more detailed description of the phonons in TBG see the App. \ref{app:phonon}. 

\begin{figure}[h]
\begin{tabular}{c}
\includegraphics[width=3.in]{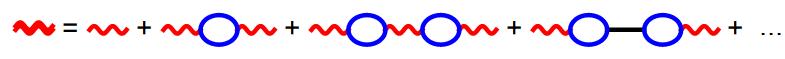} \\
\includegraphics[width=3.3in]{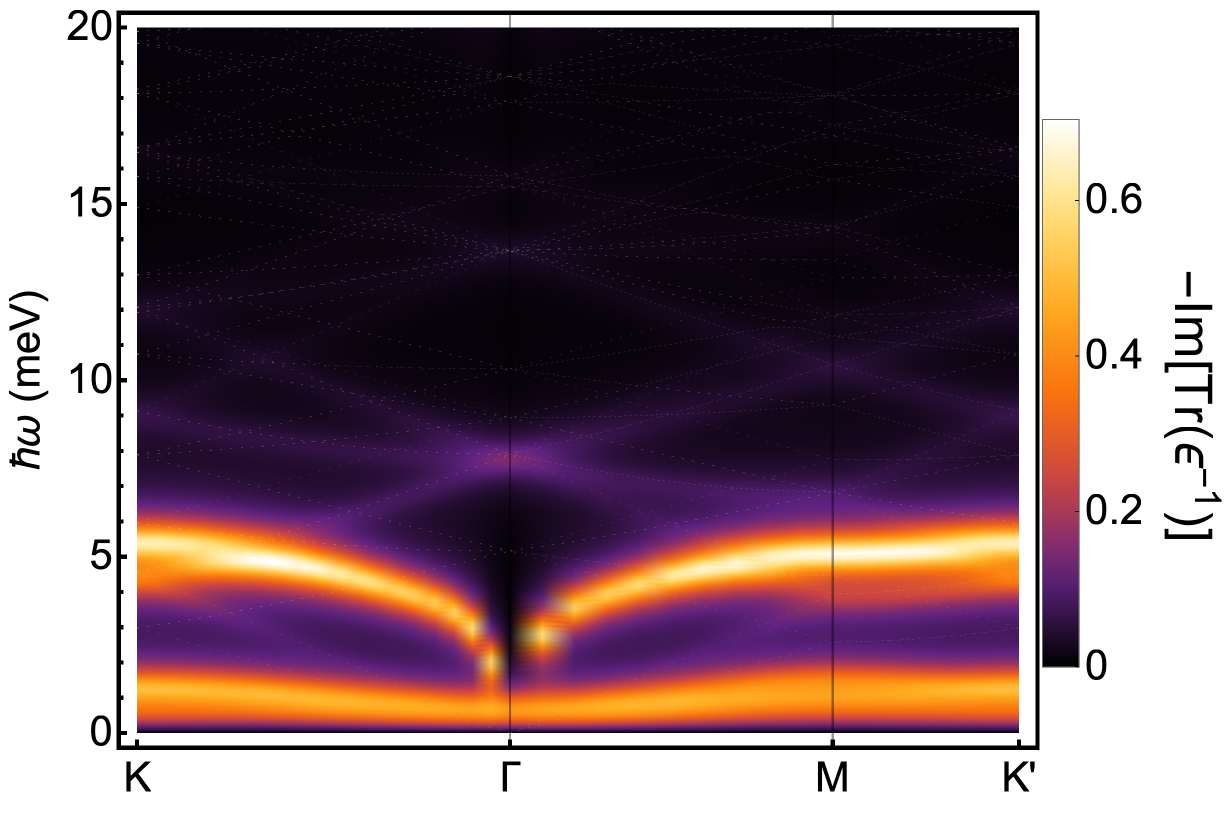}
\end{tabular}
\caption{Top: Feynman diagrams included in the screened electron-electron interaction. Red wavy lines stand for the Coulomb interaction, and straight black lines stand for the electron-phonon interaction.  Bottom: Momentum and frequency dependence of the imaginary part of the inverse dielectric function, $-{\rm Im} {\rm Tr} \, \epsilon^{-1}_{\vec{G}, \vec{G}'} ( \vec{q} , \omega )$, see \pref{inv_epsilon}. The calculation is done at the magic angle, $\theta = 1.085^\circ$, and at half filling. Due to the flatness of the bands, the system is metallic, and it has plasmons (see Sec. [\ref{sec:results}], and Fig. [\ref{fig:TCevolution}]-b)).}
\label{fig:e-h}
\end{figure}

\subsection{Polarizability}
Including the diagrams in in Fig.[\ref{fig:e-h}], top,  the full polarizability of the system can be written as:
\bea
    \chi_{\vec{G} , \vec{G}'} ( \vec{q} , \omega ) &=& \sum_{\vec{G}''} \chi^0_{\vec{G} , \vec{G}''} ( \vec{q} , \omega )  \times   \left\{ \delta_{\vec{G}'' , \vec{G}'} -  \left[ {\cal V}_C ( \vec{q} +\vec{G}'' ) +\right. \right. \nonumber\nn \\ &+&\left. \left.  {\cal V}_{eff}^{ph} ( \vec{q} + \vec{G}'' , \omega ) \right] \chi^0_{\vec{G}'' , \vec{G}'} ( \vec{q} ,\omega ) \right\}_{\vec{G}'' , \vec{G}'}^{-1} 
    \label{susc}
\eea
where $\chi^0_{\vec{G} , \vec{G}''} ( \vec{q} , \omega )$ is the bare electronic polarizability (the bubble diagram in Fig[\ref{fig:e-h}]), ${\cal V}_C(\vec{q}) = 2 \pi e^2\tanh\left(d_g|\vec{q}|\right)/(\epsilon | \vec{q} | )$ is the Coulomb potential, $d_g$ is the distance of the sample from a metallic gate, $\epsilon$ is the screening from the external environment, and the vectors $\vec{G}$'s are reciprocal lattice vectors. We use: $d_g=40$nm and $\epsilon=10$. The susceptibility in \pref{susc} is a matrix with entries labeled by reciprocal lattice vectors.

The polarization in \pref{susc} allows us to define the screened Coulomb interaction:
\begin{align}
    {\cal V}_{\vec{G} , \vec{G}'}^{scr} ( \vec{q}  , \omega ) &=  \epsilon^{-1}_{\vec{G} , \vec{G}'} ( \vec{q} , \omega ) {\cal V}_C  ( \vec{q} + \vec{G}' ),
    \label{screened}
\end{align}
with:
\begin{align}
    \epsilon_{\vec{G} , \vec{G}'} ( \vec{q} , \omega ) &=  \delta_{\vec{G} , \vec{G}'} - {\cal V}_C ( \vec{q} + \vec{G} )  \chi^{ph}_{\vec{G} , \vec{G}'} ( \vec{q} , \omega ),
    \label{inv_epsilon}
\end{align}
where $\chi^{ph}_{\vec{G} , \vec{G}'} ( \vec{q} , \omega ) $ is given by an expression similar to \pref{susc}, except that only the phonon interaction is included, in order to avoid double counting.

The diagrams which describe the screened interaction are shown in Fig. [\ref{fig:e-h}], top.
A plot of the imaginary part of the screening function, \pref{inv_epsilon}, is shown in Fig. [\ref{fig:e-h}], bottom.  
This figure measures the density of charge density excitations, weighted by their coupling to the electron quasiparticles. The lower horizontal bright bands show the electron-hole continuum, the bands directly above them are the plasmons\cite{LL19}, and the faint blue lines give the renormalized plasmons, broadened by their interaction with the electron-hole continuum. 

The analysis includes Umklapp processes which enter both in the electron-electron and in the electron-phonon interaction. The calculation in Fig. [\ref{fig:e-h}] describes as well the plasmons of the system\cite{LL19}. Further results concerning the plasmon spectrum are reported in the App. \ref{APP_plasmon_spectrum}.

\subsection{Superconducting pairing via the screened Coulomb interaction}
We consider pairing mediated by the screened interaction defined in \pref{screened}. This scheme is a variation of the method described in\cite{KL_prl65}. The polarization diagrams are iterated to infinity, using the RPA, while other, exchange like, diagrams are not considered. The use of the RPA is justified as the spin and valley degeneracy enhances the contribution of diagrams with closed electron-hole bubbles over exchange like diagrams.

The analysis of the pairing can be turned into a self-consistency condition for the propagator of a Cooper pair, or, alternatively, for the off diagonal self-energy which hybridizes electrons and holes. This self-consistency requirement reduces to an eigenvalue problem near $T_c$, where the off diagonal self-energy tends to zero. In real space and imaginary Matsubara frequencies, this linear equation is given by:
\begin{widetext}
\bea\label{BCS_vertex}
\Delta^{i_1i_2}_{\alpha\beta}(\vec{r}_1,\vec{r}_2)=
-\sum_{i \omega'} {\cal V}^{scr}(\vec{r}_1,\vec{r}_2 , i  \omega - i \omega')\int_\Omega\,d^2\vec{r}_3d^2\vec{r}_4
\sum_{i_3 i_4}K_BT
\mathcal{G}^{i_1 i_3}_{\vec{r}_1 \vec{r}_3,\alpha}\left(i\omega'\right)
\mathcal{G}^{ i_2 i_4}_{\vec{r}_2 \vec{r}_4,\beta}\left(-i\omega'\right)
\Delta^{i_3 i_4}_{\alpha\beta}(\vec{r}_3 ,\vec{r}_4),
\eea
\end{widetext}
where $\alpha , \beta$ label the spin/valley flavor,
$i_1 , i_2 , i_3 , i_4$ are sublattice and layer indices,
$\Omega$ is the area of the system,
and the $\mathcal{G}$ are the Green's functions,
which can be written as:
\begin{align}
\mathcal{G}^{ij}_{\vec{r}\vec{r}',\alpha}(i \omega) &= \sum_{n\vec{k}}\frac{\Phi^i_{n\vec{k},\alpha}(\vec{r})\Phi^{j,*}_{n\vec{k},\alpha}(\vec{r}')}{i\hbar\omega+\mu-E_{n\vec{k},\alpha}},
\end{align}
where $n$ labels the band, 
$\vec{k}$ the wave vector in the moir\'e BZ, 
$E_{n\vec{k},\alpha}$ is the band dispersion,
$\mu$ is the chemical potential, and $\Phi_{n\vec{k},\alpha}$
is the Bloch's eigenfunction corresponding to the eigenvalue $E_{n\vec{k},\alpha}$:
\begin{align}
\Phi^i_{n\vec{k},\alpha}(\vec{r})=\frac{e^{i\vec{k}\cdot\vec{r}}}{\sqrt{\Omega}}
\sum_{\vec{G}}\phi^{i}_{n\vec{k},\alpha}(\vec{G})e^{i\vec{G}\cdot\vec{r}},
\end{align}
where the $\phi$'s are eigenvectors amplitudes,
normalized according to: $\sum_{\vec{G}, i}\phi^{i,*}_{n\vec{k},\alpha}(\vec{G})\phi^{i}_{m\vec{k},\alpha}(\vec{G})=\delta_{nm}$.

We can sum over the Matsubara frequencies analytically by
considering the static limit 
of the screened potential in \pref{BCS_vertex}:
${\cal V}^{scr}(\vec{r}_1,\vec{r}_2 , i  \omega - i \omega')\simeq
{\cal V}^{scr}(\vec{r}_1,\vec{r}_2 , i\omega=0)
$. Using this approximation (to be discussed further below) we obtain the following equation for the order parameter:
\begin{align}\label{Delta_tilde_eq}
\tilde{\Delta}^{m_1m_2}_{\alpha\beta}(\vec{k})=
\sum_{n_1n_2}\sum_{\vec{q}}\Gamma^{m_1m_2}_{n_1n_2;\alpha\beta}(\vec{k},\vec{q})
\tilde{\Delta}^{n_1n_2}_{\alpha\beta}(\vec{q}),
\end{align}
where $\tilde{\Delta}^{m_1m_2}_{\alpha\beta}(\vec{k})$
defines the amplitude for the pairing between the bands $m_1$ and $m_2$, and the kernel $\Gamma$ is given by:
\begin{widetext}
\bea
\Gamma^{m_1m_2}_{n_1n_2;\alpha\beta}(\vec{k},\vec{q}) &=&
-\frac{1}{\Omega}\sum_{\vec{G_1}\vec{G_1}'}\sum_{\vec{G_2}\vec{G_2}'}\sum_{i_1i_2}
{\cal V}^{scr}_{\vec{G}_1-\vec{G}_1',\vec{G}_2-\vec{G}_2'}
\left(\vec{k}-\vec{q}\right)\times 
 \phi^{i_1,*}_{m_1\vec{k},\alpha}(\vec{G}_1)
\phi^{i_2,*}_{m_2-\vec{k},\beta}\left(\vec{G}_2'\right)
\phi^{i_1}_{n_1\vec{q},\alpha}\left(\vec{G}_1'\right)
\phi^{i_2}_{n_2-\vec{q},\beta}(\vec{G}_2)\times\nn\\
&\times&
\sqrt{
\frac{
f\left(-E_{m_2-\vec{k},\beta}+\mu\right)-f\left(E_{m_1\vec{k},\alpha}-\mu\right)
}{E_{m_2-\vec{k},\beta}+E_{m_1\vec{k},\alpha}-2\mu}}\times 
\sqrt{
\frac{
f\left(-E_{n_2-\vec{q},\beta}+\mu\right)-f\left(E_{n_1\vec{q},\alpha}-\mu\right)
}{E_{n_2-\vec{q},\beta}+E_{n_1\vec{q},\alpha}-2\mu}},
\label{Gamma_eq}
\eea
\end{widetext}
where $f ( E )$ is the Fermi-Dirac distribution function.
The detailed derivation of the Eqs. (\ref{Delta_tilde_eq}) and (\ref{Gamma_eq}) is given in the App. \ref{APP_KL_formalism}.

The condition for the onset of superconductivity is that the kernel $\Gamma$ has an eigenvalue equal to 1.
This defines the critical temperature, $T_c$, as the one at which the largest eigenvalue of $\Gamma$ is equal to 1.

As the Hamiltonian of the TBG does not depend on the spin,
we can identify two different kinds of vertices,
depending on whether $\alpha$ and $\beta$ share the same or opposite valley indices.
These two vertices describe intra-valley or inter-valley superconductivity.
As we argue in the App. \ref{APP_KL_formalism},
the superconductivity is generally favored in the inter-valley channel,
and consequently we focus on that case in the following study.

Note, finally, that the potential ${\cal V}^{scr}_{\vec{G} , \vec{G}'} ( \vec{q} )$ is repulsive. The attractive phonon induced electron-electron interaction enters only in the screening function, and it is smaller than the Coulomb potential, except for Umklapp processes characterized by large reciprocal lattice vectors, $\vec{G}$. These terms are suppressed by the low electron polarizability at these wavevectors.

We neglect pairing due to the bare electron-phonon coupling. 
The dimensionless constant which describes approximately the phonon induced electron-electron interaction, $\overline{g} \approx 0.4$ implies weak-medium interaction, partially because only phonons of wavelength comparable to the moir\'e period can induce pairing.

A simple approximation, which allows us to perform analytically the inversion of the dielectric matrix, and to isolate diagonal and off diagonal Umklapp terms is given in the App. \ref{App_anal_Vscr}.
\\
\section{Results}
\label{sec:results}
\begin{figure}
\begin{tabular}{c}
\includegraphics[width=2.5in]{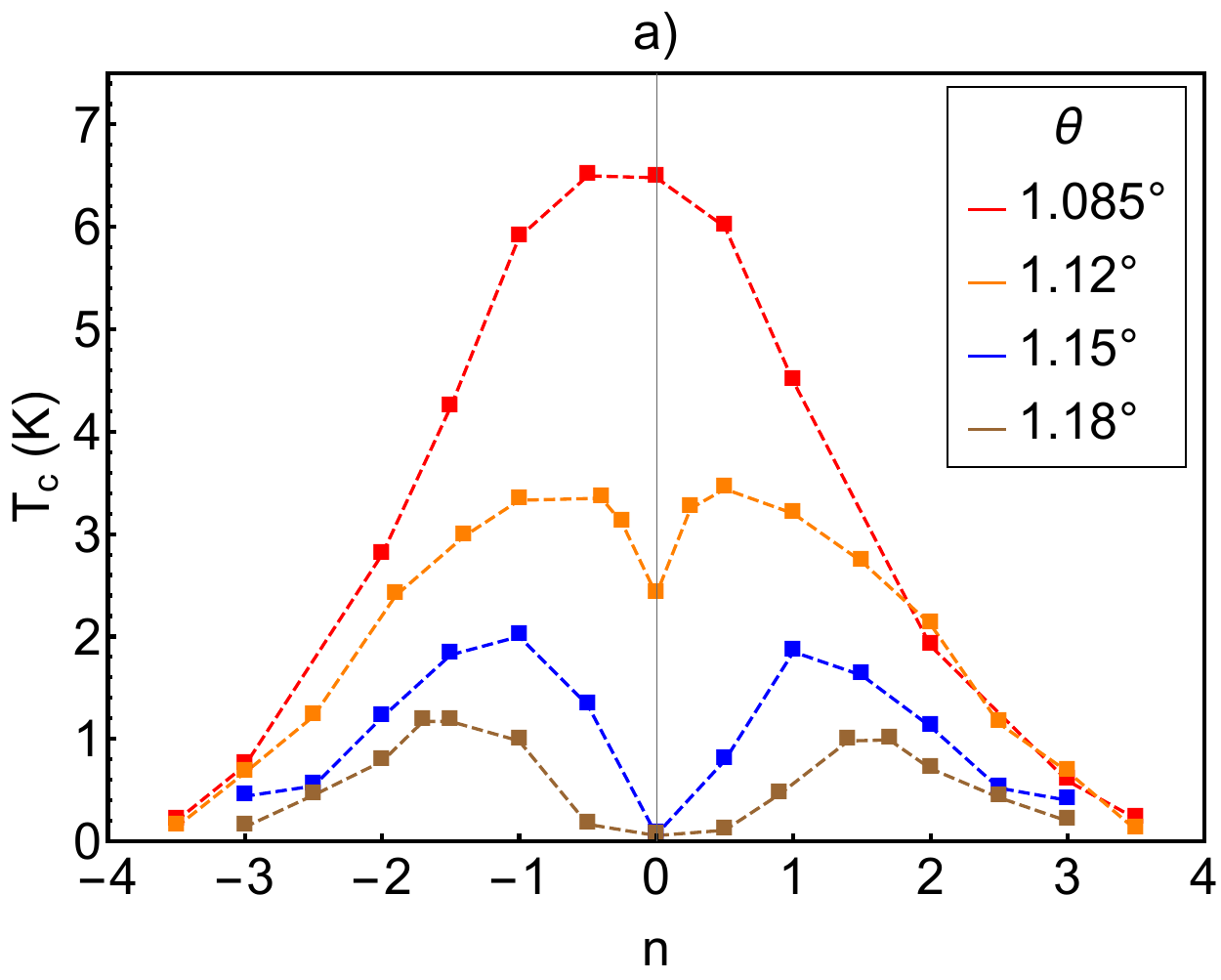} \\
\includegraphics[width=2.5in]{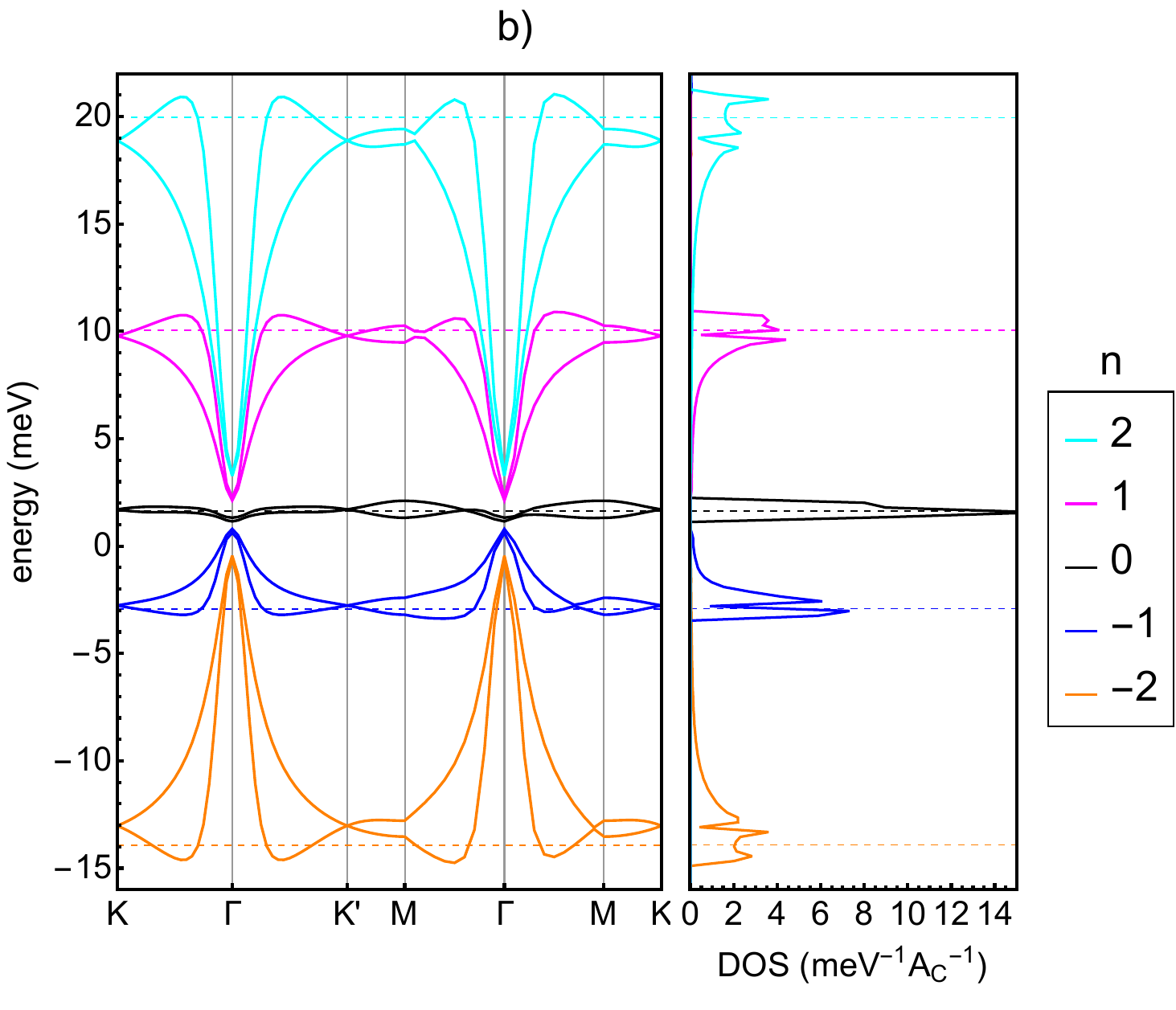}
\end{tabular}
\caption{a) values of the critical temperature as a function of filling for various angles (in our definition, $-4 \le n \le + 4$, and $n=0$ denotes the charge neutrality point). b) bands and DOS for the magical angle $\theta = 1.085^\circ$, and different fillings. The horizontal lines mark the position of the Fermi energy.}
\label{fig:TCevolution}
\end{figure}

Results for the critical temperature, for different angles and fillings   are shown in Fig. [\ref{fig:TCevolution}]-a). The calculations take into account the Hartree potential, but not the exchange term. Note that, at the magic angle, $\theta = 1.085^\circ$, and at half filling, when the Hartree potential vanishes, the systems is metallic, with three pockets at the Fermi surface, see  Fig. [\ref{fig:TCevolution}]-b). 
Note that the exchange potential, not considered here, leads to a gap at half filling\cite{CG20,ZJVZ20,LKYV21}. This gap is of order $\sim 20 -40$ meV, that is, much larger than the value of $T_c$ calculated here. Hence, it seems likely that the insulating phase will prevail at half filling and  close to a magic angle.

The value of $T_c$ approximately follows the DOS at the Fermi level. As function of filling, the regions near van Hove singularities, which are displaced by the Hartree potential\cite{cea_prb19}, give the maxima of $T_c$.

\begin{figure}[h]
\includegraphics[width=3in]{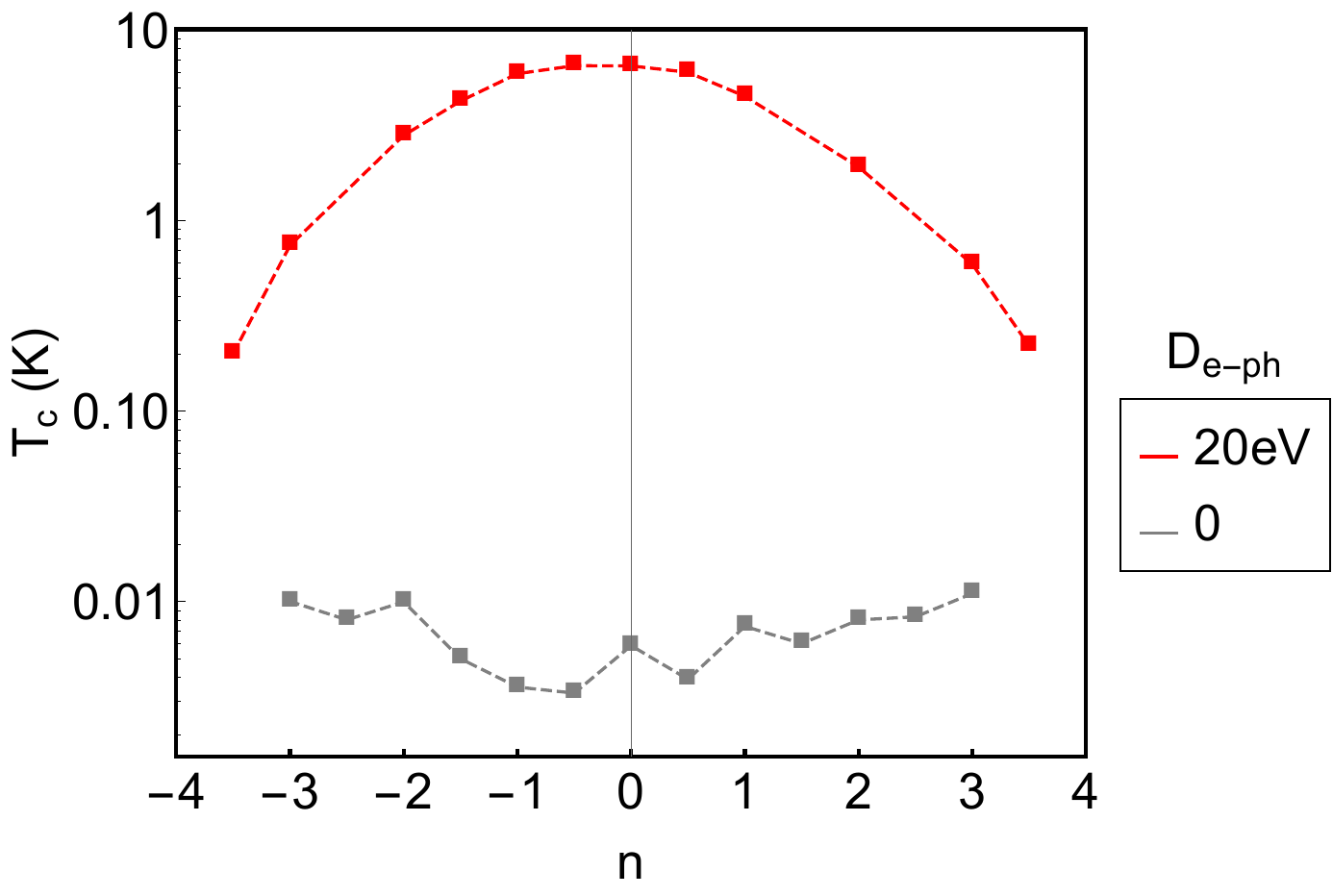}
\caption{Superconducting critical temperature with (red curve) and without (black curve) the electron-phonon contribution to the screening of the interaction. The calculation is at the magic angle, $\theta = 1.085^\circ$. Note the logarithmic scale.}
\label{fig:comparison}
\end{figure}

The results of Fig. [\ref{fig:comparison}] show that TBG is superconducting also in the absence of electron-phonon coupling.
The critical temperature is of order $T_c \sim 10^{-2}$ K. Similar results, although with somewhat smaller values for the critical temperature, have been obtained in\cite{KZB20}.

A discussion about the effect of the external screening on superconductivity is provided in the App. \ref{App_screening}. Remarkably, our results predict that superconductivity becomes more robust upon increasing the external screening, in good agreement with recent experimental findings\cite{Liu_science2021}. 

In order to check if the superconductivity found here is related to the complexity of the wavefunctions of TBG (see, for instance\cite{PWV18,PZSV19,SWSLFB19,HLSCB19,PM20}), or it is a consequence of an enhanced DOS, we have performed calculations using the parameters in Fig. [\ref{fig:TCevolution}], including the electron-phonon coupling, neglecting the contribution from Umklapp processes. We obtain very low critical temperatures, of order $T_c \sim 10^{-4}$K.

A comprehensive analysis of the role of Umklapp processes in the pairing is beyond the scope of this work. An estimate of their relevance can be inferred from the lowest order (zero and one bubble diagrams) in Fig.[\ref{fig:e-h}]. Let us assume that the number of relevant reciprocal lattice vectors $\vec{G}$ is $n_G$ (including $\vec{G} = 0$). For a momentum transferred $\vec{q}$, there are $n_G$ direct interactions, defined by the bare interaction $v^C_{\vec{G} + \vec{q}}$. There are $n_G^2$ second order diagrams, of order $v^C_{\vec{G} + \vec{q}} v^C_{\vec{G}' + \vec{q}} \chi_{\vec{G}, \vec{G}'} ( \vec{q} )$. These diagrams are attractive, as $\chi_{\vec{G}, \vec{G}'} ( \vec{q} ) < 0$. Hence, the combination of first and second order processes lead to an attractive interaction for $n_G \gg 1$ and $| \vec{q} | \sim \overline{|\vec{G}|}$.

The relevance of Umklapp processes shows that, besides a large DOS, the non trivial nature of the wave functions is crucial for the appearance of superconductivity\cite{metric}.

\begin{figure}[h]
\begin{tabular}{c}
\includegraphics[width=3.5in]{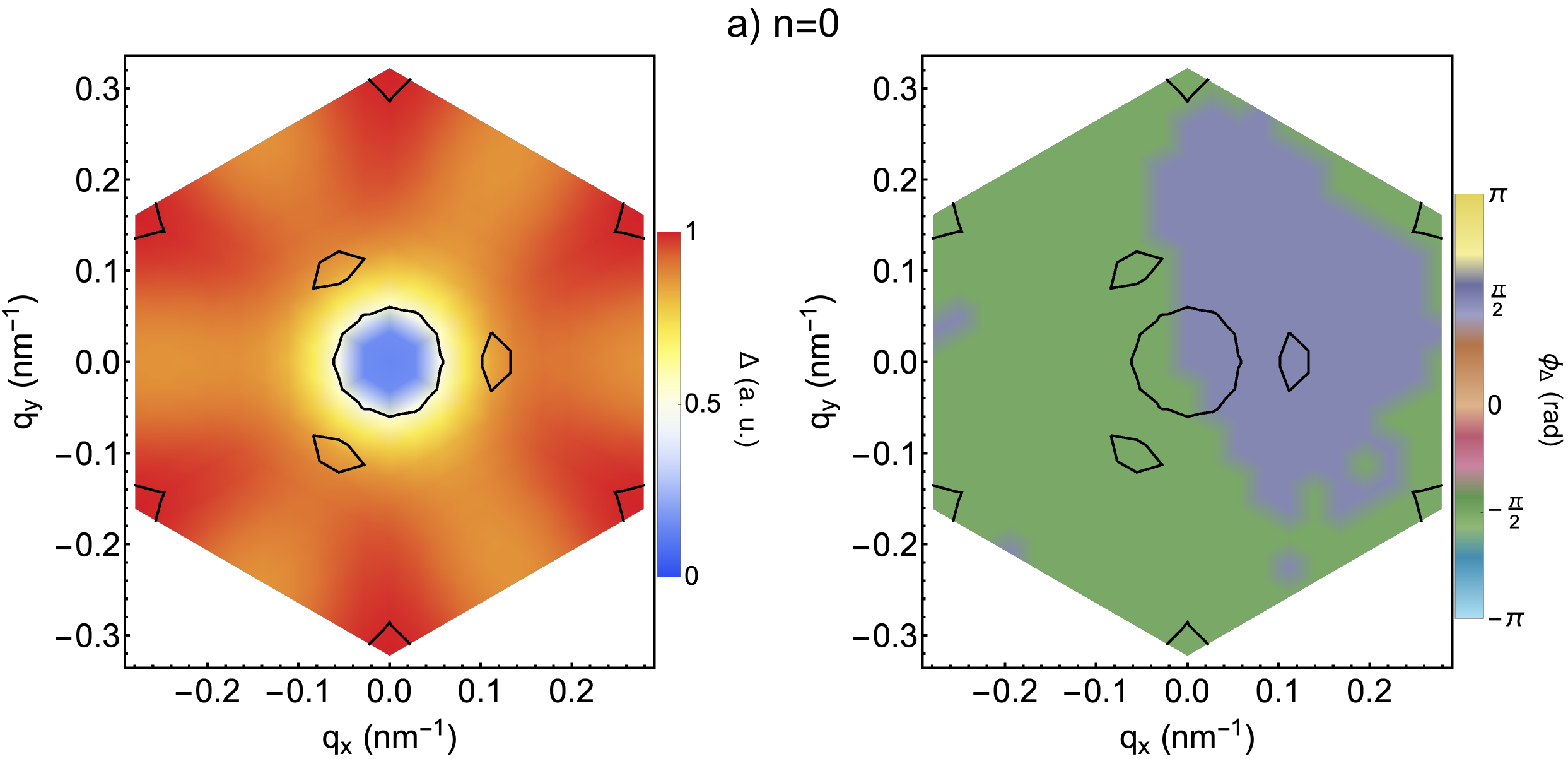} \\
\includegraphics[width=3.5in]{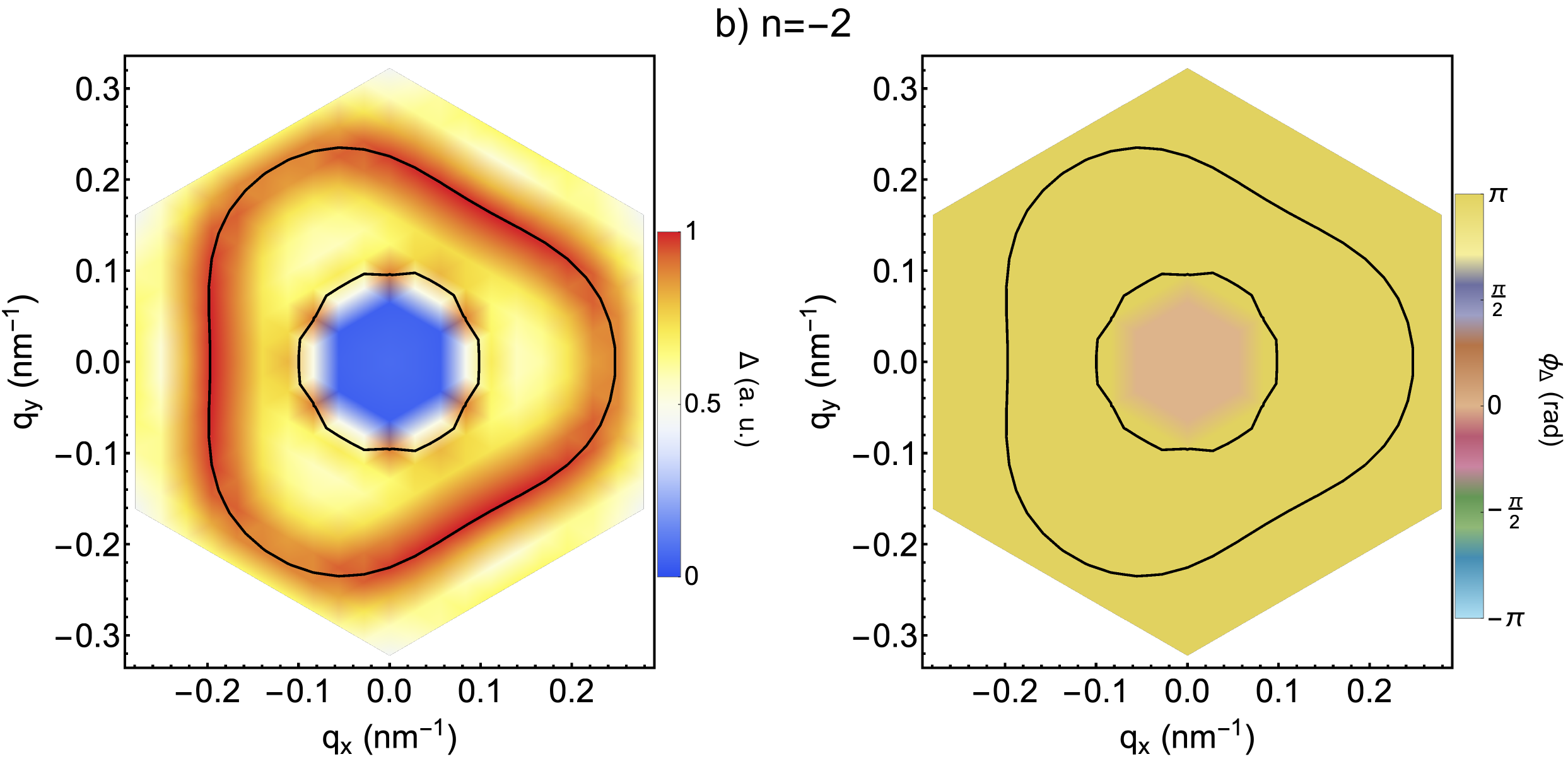}
\end{tabular}
\caption{Absolute value (left) and phase (right) of the order parameter projected onto the valence band throughout the BZ at the magic angle, $\theta = 1.085^\circ$, and fillings: $n = 0$ (a) and $n = -2$ (b). The black lines show the Fermi surface.}
\label{fig:OP_symmetry}
\end{figure}

The absolute value and the phase of the order parameter projected onto the valence band throughout the BZ are shown in Fig. [\ref{fig:OP_symmetry}]. For a filling $n = 0$ the width of the pair of bands is $\lesssim 1$ meV, and all the states in the central bands contribute to the pairing. The phase of the order parameter shows a change of sign, although lack of numerical accuracy prevents a precise determination of the symmetry, which seems close to p-wave. For $n = -2$, the order parameter is mostly localized in the vicinity of the two Fermi surfaces. The system resembles a two gap superconductor, with generalized $s$-wave symmetry.

The number of pockets, and detailed shape of the Fermi surface is highly dependent on the parameters of the model. A robust feature of our results, however, is the lack of sign changes of the order parameter along the Fermi surfaces, irrespective of their shape. A purely repulsive interaction favors sign changes, as in the p-wave pairing in $^3$He or the d-wave pairing in the cuprates. We ascribe this effect to the fact that the screened interaction shows sizable attractive regions, as mentioned previously.

The calculations leading to the order parameter shown in Fig. [\ref{fig:OP_symmetry}] have been done assuming pairing between an electron in one valley with momentum $\vec{k}$ and another electron in the other valley, with momentum $- \vec{k}$ (see SI, section 4). Hence, the order parameter is compatible with spin singlet and spin triplet pairing, as the antisymmetry is ensured by using the right combination of the spin and valley components\cite{SS20,KLV20}.

Further results about the symmetry of the order parameter are reported in the App. \ref{App_OP_symmetry}. 

The approximate analysis of the pairing potential in App. \ref{App_anal_Vscr} suggests that Umklapp processes favor pairing near the edges of the BZ, and reduce it near its center, the $\Gamma$ point. The role of Umklapp processes is further analyzed in the App. \ref{APP_pairing_potential}. 

\section{Discussion}
\subsection{Main results}
\label{sec:results}
The work presented here addresses two main topics:
\begin{itemize}
    \item 
An analysis of the excitations of TBG. The calculations describe electron-hole pairs, plasmons, and acoustic phonons. 
For simplicity, we have only included longitudinal phonons, although the calculations can easily be extended to transverse acoustic\cite{LWB19} and to optical phonons\cite{WMM18}. Intravalley phonons can be expected to increase further the superconducting pairing. The effect of inter-valley optical phonons\cite{WMM18,ATF19} will enhance or suppress the pairing depending on whether the paired electrons are in a spin singlet/valley triplet or in a spin triplet/valley singlet combination.

 The results reported here lead to the screening of the deformation potential, and to the renormalization of both the electron-hole pairs, and the phonon dispersion. The electron bands include the effects of the Hartree potential, but not exchange contributions.

We find a significant renormalization of the electron-hole continuum due to plasmons and phonons, and reciprocal changes in the phonon dispersion.

\item
We have analyzed superconducting pairing due to the repulsive Coulomb interaction screened by the excitations mentioned above. We do not consider here the direct, attractive, electron-phonon interaction, as it can be expected to give a small correction. Superconductivity is closely linked to the contribution of Umklapp processes.

Near half filling, where the central bands are narrowest, all states in the BZ contribute to the order parameter. Away from half filling, the weight of the order parameter is concentrated near the Fermi surface.

In most cases, we find that the order parameter has different weights in different pockets of the Fermi surface. The system resembles a multigap superconductor, with extended $s$-wave symmetry. The lack of sign changes in the order parameter, common in superconductivity derived from repulsive interactions, is due to the fact that Umklapp processes induce attractive terms in the superconducting kernel, \pref{Gamma_eq} (see also the App. \ref{APP_pairing_potential}). We have studied Cooper pairs made from electrons in different valleys. The resulting order parameter is compatible with spin singlet and with spin triplet pairing.

\end{itemize}

\subsection{Other effects}
\label{sec:effects}
The work does not attempt to give numerically accurate predictions of the superconducting critical temperature, but the analysis is expected to give reasonable order of magnitude estimates, and to identify trends. 

By considering together the role of electron-hole excitations, plasmons, and phonons in the dielectric function, we obtain values of $T_c$ of the same order of magnitude as the experimentally measured ones. The value of $T_c$ follows, approximately, the DOS at the Fermi surface. The suppression of the phonon contribution to the screening reduces the values of $T_c$ by about two orders of magnitude. The neglect of the complexity of the wavefunctions, by leaving out Umklapp processes, reduces $T_c$ by one or two orders of magnitude more, even including phonons. On the other hand, the order of magnitude of the critical temperature does not change when only the central bands are included in the calculation.

Our analysis does not consider exchange effects. In a fully symmmetric, metallic phase, the exchange term is significantly smaller than the Hartree potential. The exchange contribution can lead to broken symmetry phases, where our analysis does not apply. Near a filling $n = \pm 2$, however, Hartree Fock results\cite{CG20}, and the "cascade" picture of multiple phase transitions\cite{Zetal20}, lead to two completely full pairs of bands, and two pairs of partially filled bands. If these bands belong to opposite valleys, our analysis, based on inter-valley pairing, should be qualitatively correct.

For simplicity, we have not included the effect of transverse phonons, which couple to the electrons via a gauge potential\cite{NGPNG09,VKG10} (see also the App. \ref{app:phonon}). The inclusion of these phonons will increase the tendency towards superconductivity.
The frequency dependence of the phonon propagator, not considered here, leads to an upper bound in the value of $T_c$, which cannot exceed the phonon frequencies.

Elastic scattering is expected to induce pair breaking, as the order parameter has different values at different pockets. These effects will change substantially the superconducting properties when the elastic mean free path is comparable to the inverse of the separation between pockets, ${\ell}_{el} \sim | \Delta \vec{k} |^{-1}$. This estimate gives a mean free path a few times larger than the moir\'e wavelength. If the order parameter has opposite signs in the two valleys, defects which induce inter-valley scattering will lead to subgap Andreev states.

\section{Acknowledgements}
\label{sec:acknowledgements}
{We are thankful to Leonid Levitov, Liang Fu and José Maria Pizarro for many productive conversations.
This work was supported by funding from the European Commision, under the Graphene Flagship, Core 3, grant number 881603, and by the grants NMAT2D (Comunidad de Madrid, Spain),  SprQuMat and SEV-2016-0686, (Ministerio de Ciencia e Innovación, Spain).


\appendix
\def\appendixname{Appendix}
\section{The continuum model of TBG}\label{APP_continuum_model}
We describe the TBG within the low energy continuum model
considered in  Refs.\cite{LPN07,BM11,koshino_prx18},
which is meaningful for sufficiently small angles, $\theta$,
so that an approximatively commensurate structure can be defined for any twist.
The period of the moir\'e is: $L=\frac{a}{2\sin(\theta/2)}$,
$a=2.46${\AA} being the lattice constant of graphene.
The moir\'e BZ, resulting from the folding of the two BZs of each monolayer,
is spanned by the two reciprocal lattice vectors:
\begin{equation}
\vec{G}_1=2\pi(1/\sqrt{3},1)/L\text{ and } \vec{G}_2=4\pi(-1/\sqrt{3},0)/L.
\end{equation}
For small $\theta$, one can safely neglect the coupling between the two valleys of the monolayer graphene at $K=4\pi(1,0)/(3a)$ and $K'=-K$.
For each valley,
the Hamiltonian of the TBG is a $4\times4$ matrix, with entries: $\left(A_b,B_b,A_t,B_t\right)$,
where $A,B$ denote the sub-lattice and $b,t$ refer to the bottom and top layer, respectively.
Without loss of generality, we assume that in the aligned configuration, at $\theta=0^\circ$,
the two layers are $AA$-stacked.
The continuum Hamiltonian of the TBG can be generally written as\cite{LPN07,BM11,koshino_prx18}:
\bea\lb{HTBG}
H_K=
\begin{pmatrix}
H_b&U(\vec{r})\\U^\dagger(\vec{r})&H_t
\end{pmatrix}\quad,\quad
H_{K'}=H_K^*\quad,
\eea
where:
\bea
H_l=\hbar v_F \left(-i\vec{\nabla}-K_l\right)\cdot
\vec{\tau}_{\theta_l}
\eea
is the Dirac Hamiltonian for the $K$-valley of the layer $l=b,t$, $v_F=\sqrt{3}ta/(2\hbar)$ is the Fermi velocity, $t$ is the hopping amplitude between localized $p_z$ orbitals at nearest neighbors carbon atoms, $\theta_{b,t}=\mp\theta/2$, $\vec{\tau}_{\theta_l}=e^{i\tau_z\theta_l/2}\left(\tau_x,\tau_y\right)e^{-i\tau_z\theta_l/2}$,
$\tau_i$ are the Pauli matrices, and $K_l=4\pi\left(\cos\theta_l,\sin\theta_l\right)/(3a)$
are the Dirac points of the two twisted monolayers, which identify the corners of the moir\'e BZ. $U(\vec{r})$ is the interlayer potential, which is a periodic function in the moir\'e unit cell. In the limit of small angles, its leading harmonic expansion is determined by only three reciprocal lattice vectors\cite{LPN07}:
$U(\vec{r})=U(0)+U\left(-\vec{G}_1\right)e^{-i\vec{G}_1\cdot\vec{r}}+
U\left(-\vec{G}_1-\vec{G}_2\right)e^{-i\left(\vec{G}_1+\vec{G}_2\right)\cdot\vec{r}}$,
where the amplitudes $U\left(\vec{G}\right)$ are given by:
\begin{widetext}
\bea
U(0)=\begin{pmatrix}g_1&g_2\\g_2&g_1\end{pmatrix}\quad,\quad
U\left(-\vec{G}_1\right)=\begin{pmatrix}g_1&g_2e^{-2i\pi/3}\\g_2e^{2i\pi/3}&g_1\end{pmatrix}\quad,\quad
U\left(-\vec{G}_1-\vec{G}_2\right)=\begin{pmatrix}g_1&g_2e^{2i\pi/3}\\g_2e^{-2i\pi/3}&g_1\end{pmatrix}.
\eea
\end{widetext}
In the following we adopt the parametrization of the TBG given in the Ref.\cite{koshino_prx18}: $\hbar v_F/a=2.1354$eV, $g_1=0.0797$eV and $g_2=0.0975$eV. The difference between $g_1$ and $g_2$, as described in\cite{koshino_prx18}, accounts for the inhomogeneous interlayer distance, which is minimum in the $AB/BA$ regions and maximum in the $AA$ ones, or it can be seen as a model
of a more detailed treatment of lattice relaxation\cite{Guinea2019}. These parameters are consistent with atomic calculations of the lattice relaxation\cite{LDCNC19,CACCNL20}.

The Hamiltonian of the Eq. \pref{HTBG} corresponding to the valley $\xi K$,
with $\xi=\pm$, is diagonalized by Bloch eigenfunctions satisfying periodic boundary conditions in a region of space, $\Omega$, containing a large number of moir\'e unit cells:

\bea\label{Bloch_waves}
\Phi^i_{n\vec{k},\xi}(\vec{r})=\frac{e^{i\vec{k}\cdot\vec{r}}}{\sqrt{\Omega}}
\sum_{\vec{G}}\phi^{i}_{n\vec{k},\xi}(\vec{G})e^{i\vec{G}\cdot\vec{r}},
\eea
where $n$ is the band index, $\vec{k}$ is the wave vector in the moir\'e BZ,
$\vec{G}=n_1\vec{G}_1+n_2\vec{G}_2$ are reciprocal lattice vectors, with $n_1,n_2$ integers,
$i$ is the sub-lattice/layer index
and $\phi^{i}_{n\vec{k},\xi}$ are numerical eigenvectors,
normalized according to: $\sum_{\vec{G}i}\phi^{i,*}_{n\vec{k},\xi}(\vec{G})\phi^{i}_{m\vec{k},\xi}(\vec{G})=\delta_{nm}$.
Upon varying $\vec{k}$, the eigenvalue $E_{n\vec{k},\xi}$ corresponding to $\Phi_{n\vec{k},\xi}$ defines the $n$-th Bloch band of the $\xi$-valley.
The eigenfunctions and energies in opposite valleys are related each other by a combined conjugate and time-reversal symmetry:
$\{\Phi^i_{n\vec{k},-}(\vec{r}), E_{n\vec{k},-} \}=
\{\Phi^{i,*}_{n-\vec{k},+}(\vec{r}), E_{n-\vec{k},+} \}$.

In the numerics, the number of Fourier components defining the eigenfunctions in the Eq. \pref{Bloch_waves} is bounded by a cutoff: $|\vec{G}|\le G_c$, where $G_c$ is chosen in order to achieve the convergence of the low energy bands.
In the present, we use $G_c=5\left|\vec{G}_1\right|$, which is sufficient to attain the convergence within the relevant window of a few meV's from the charge neutrality point.
    \\
\section{Plasmon spectrum in TBG}\label{APP_plasmon_spectrum}
The Fig. \ref{fig:plasmon_spectrum} shows the spectral function of the plasmon as a function of the frequency and the wave vector, and fillings: $n=0(a),-1(b),-2(c)$.
The plasmon dispersion displays the typical $\sim\sqrt{k}$ behavior characterising the surface plasmons. In addition, the plasmon coexists with both the phonons and the particle-hole continuum associated to the intra- and inter-band transitions between the two central bands. Interestingly, upon moving the filling from the charge neutrality point, $n=0$, the particle-hole continuum moves towards higher energies, according to the evolution of the bandwidth, and it overcomes the plasmon line for $n\lesssim-2$. 
\begin{figure}[h!]
\centering
\includegraphics[width=3.in]{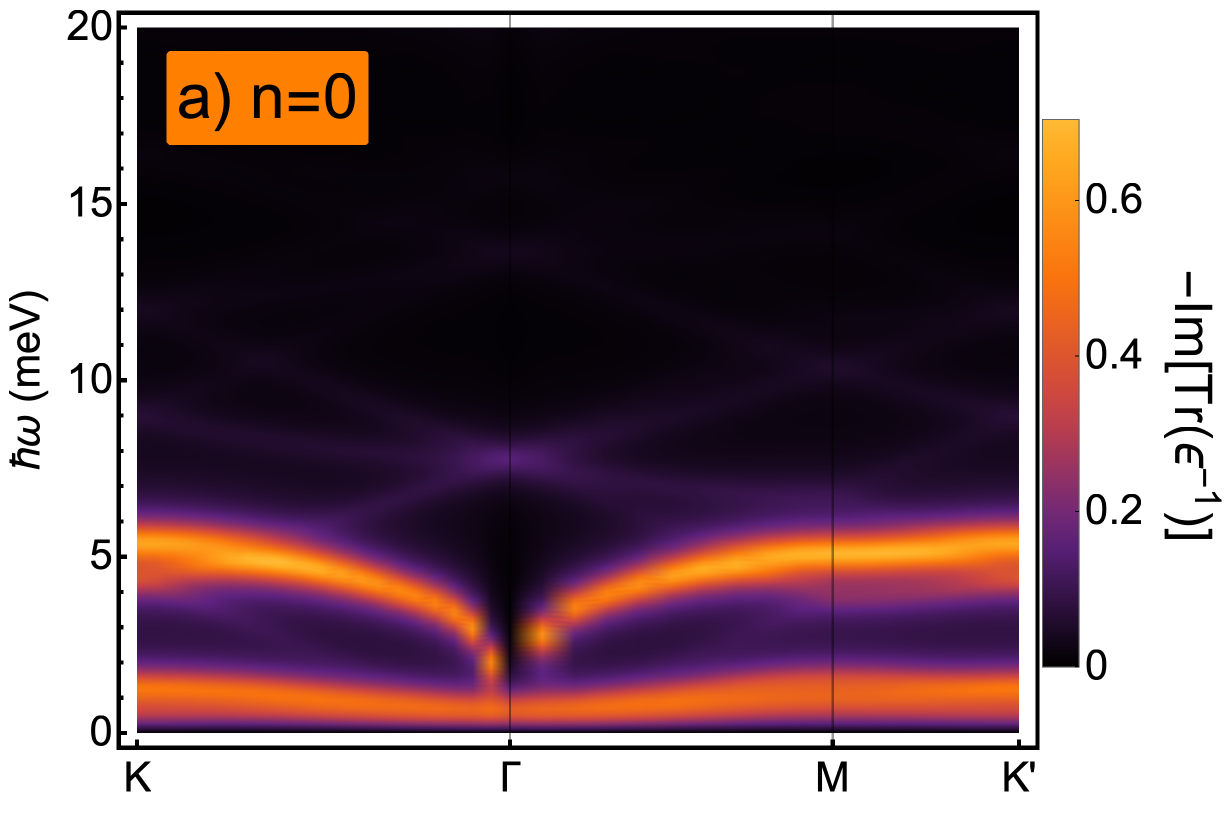}\\
\includegraphics[width=3.in]{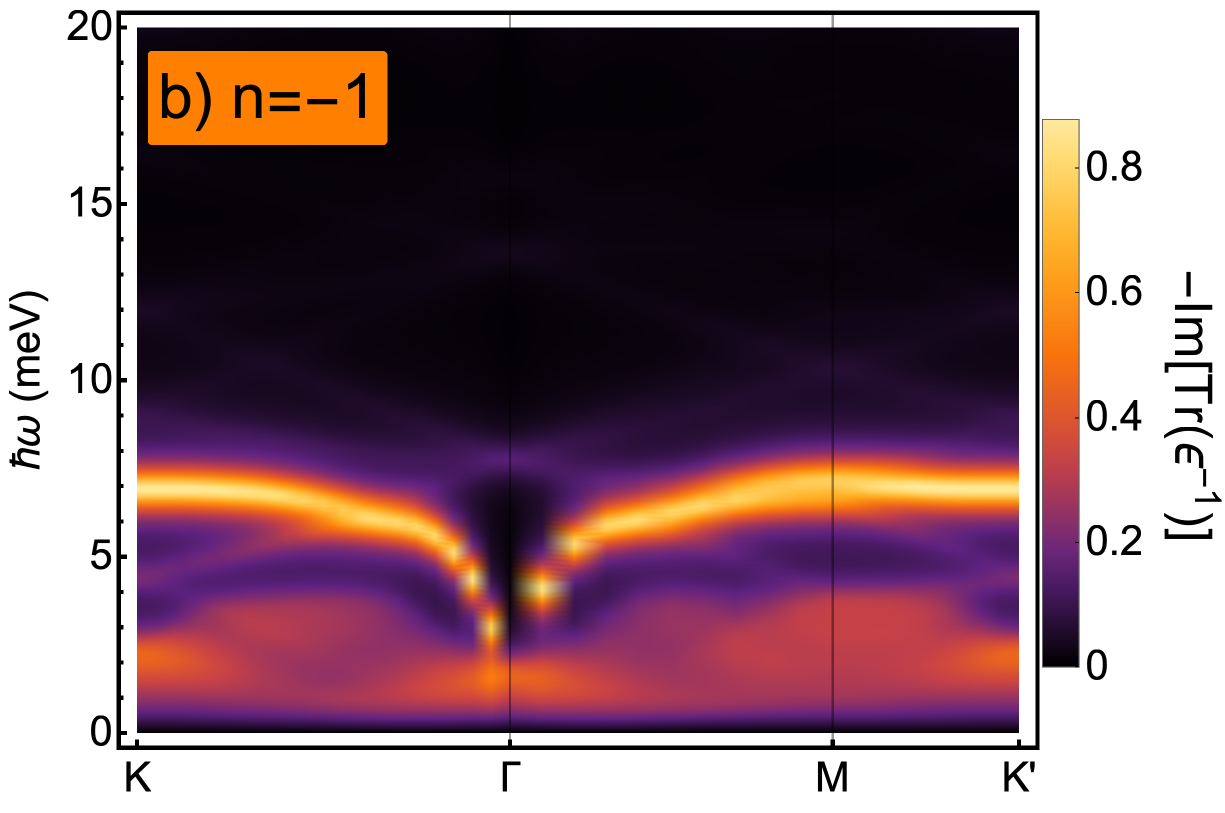}\\
\includegraphics[width=3.in]{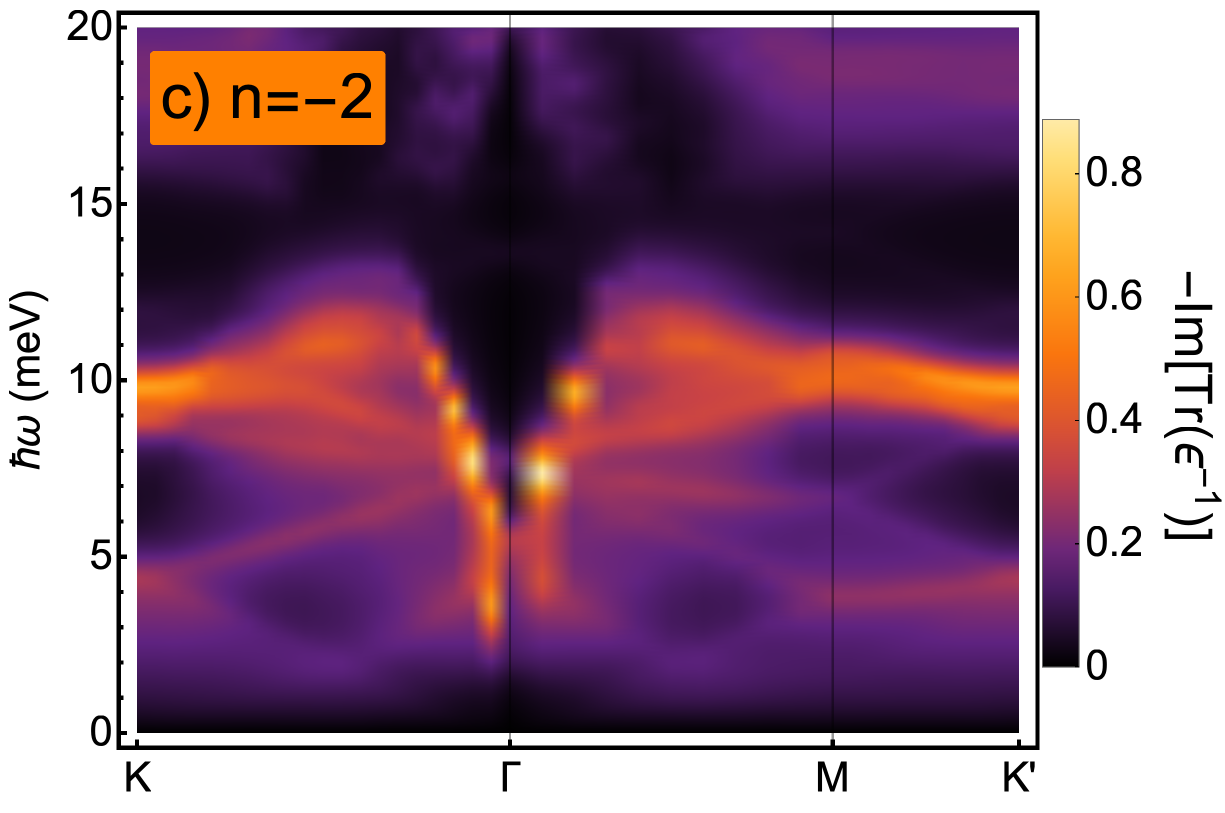}
\caption{
Plasmon spectral function as a function of the frequency and the wave vector, and fillings: $n=0(a),-1(b),-2(c)$.
}
\label{fig:plasmon_spectrum}
\end{figure}
Cuts of the plasmon spectral function obtained at the wave vectors: $q=K$ and $q=M$, are shown in the Fig. \ref{fig:plasmon_spectrum_cuts}. Note that the density of low energy modes below the phonon frequencies is distorted, and changed with respect the Fermi liquid behavior ${\rm Im} \epsilon^{-1} ( \vec{q} , \omega ) \propto | \omega |$. These changes will influence the magnitude and temperature dependence of the inelastic scattering time in the normal phase\cite{WHS19,Petal19b,Cetal19,Yetal19,GS20}.
\begin{figure}[h!]
\centering
\includegraphics[width=2.5in]{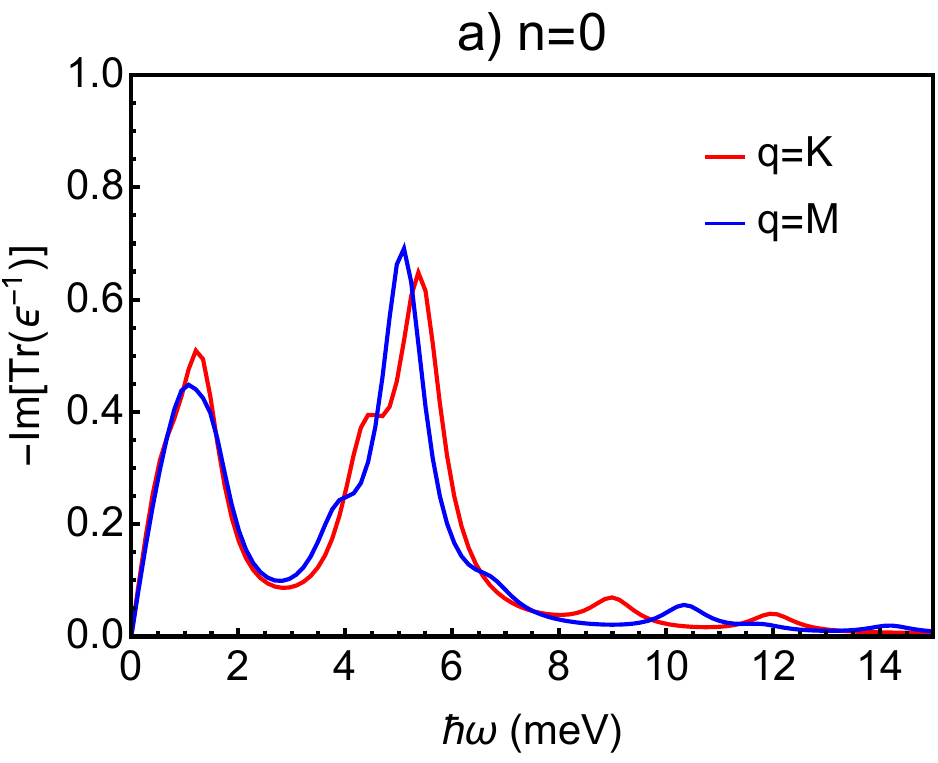}\\
\includegraphics[width=2.5in]{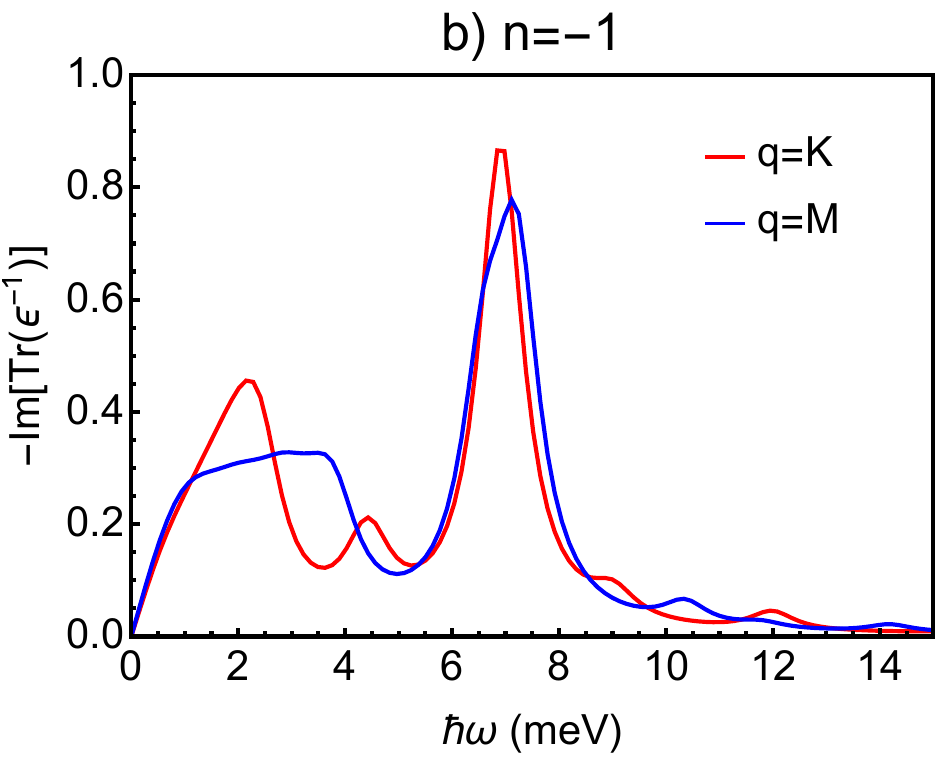}\\
\includegraphics[width=2.5in]{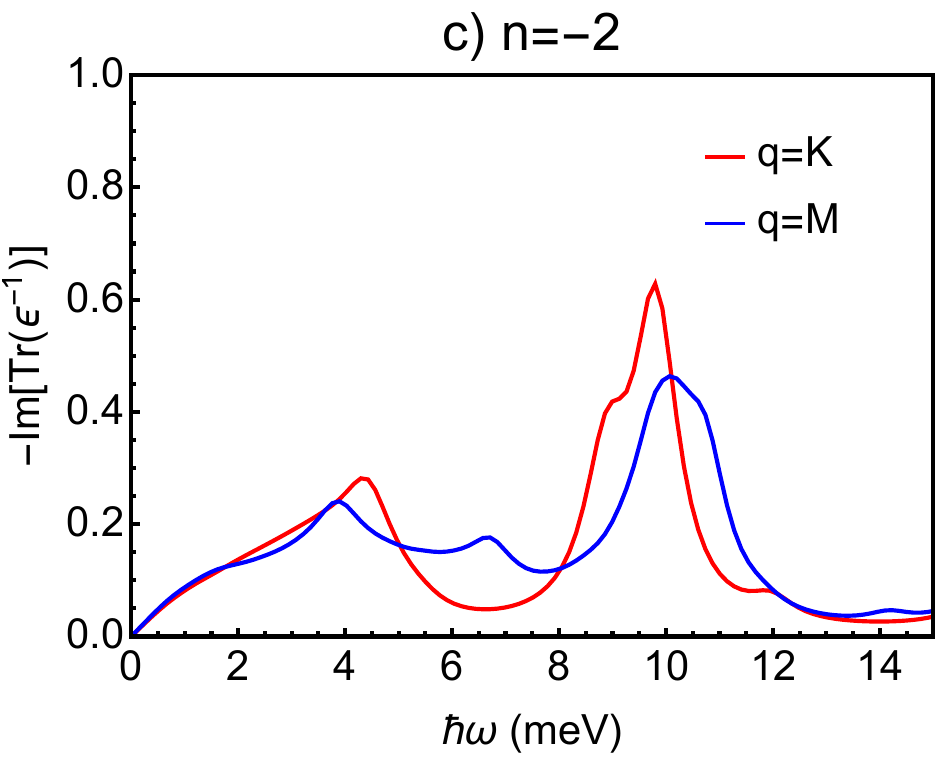}
\caption{
Cuts of the plasmon spectral function of Fig. \ref{fig:plasmon_spectrum} at the wave vectors: $q=K$(red line) and $q=M$(blue line).
}
\label{fig:plasmon_spectrum_cuts}
\end{figure}

The plasmon spectral functions obtained without including the phonon contribution, i.e. by neglecting the electron-phonon coupling: $D=0$ 
eV, are shown in the Fig. \ref{fig:plasmon_spectrum_D=0eV}.
Upon comparing the Figs. \ref{fig:plasmon_spectrum} and \ref{fig:plasmon_spectrum_D=0eV} we can note that the contribution of the phonon does not alter drastically the plasmon dispersion. 
\begin{figure}[h!]
\centering
\includegraphics[width=3.in]{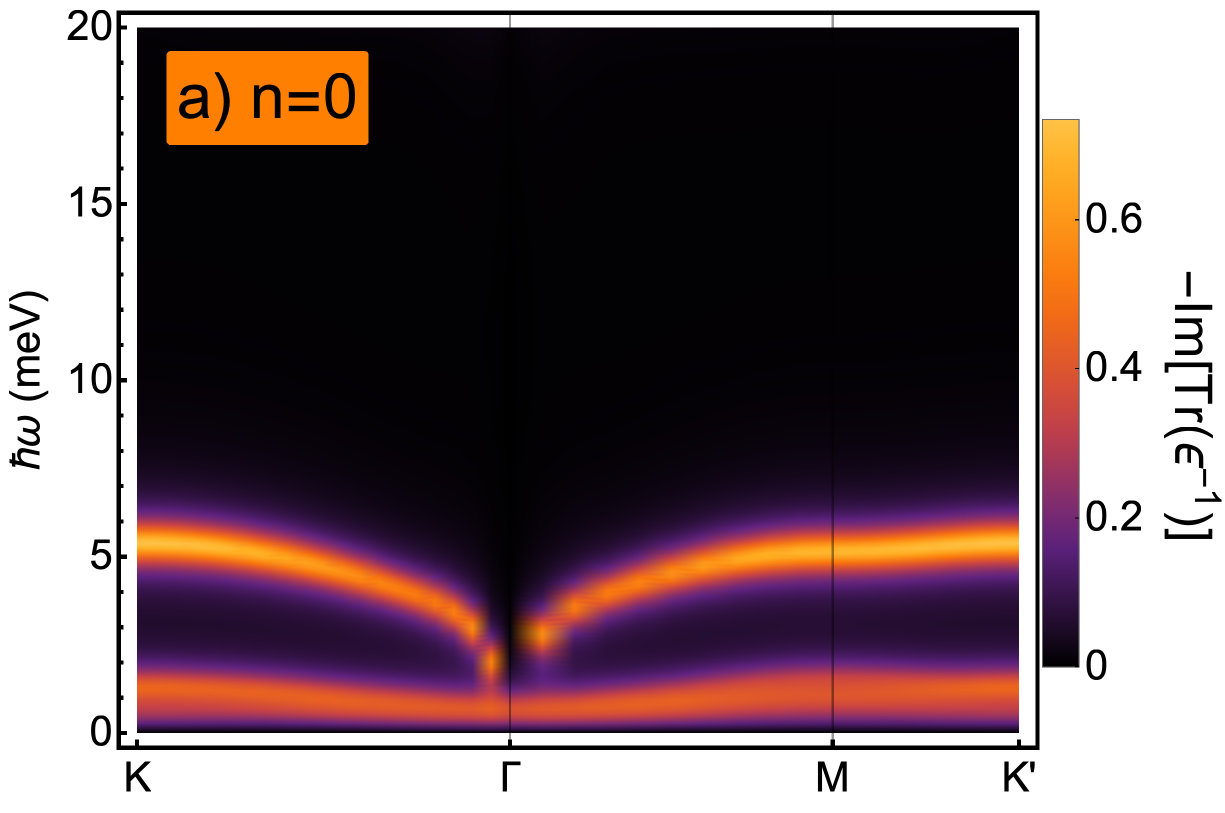}\\
\includegraphics[width=3.in]{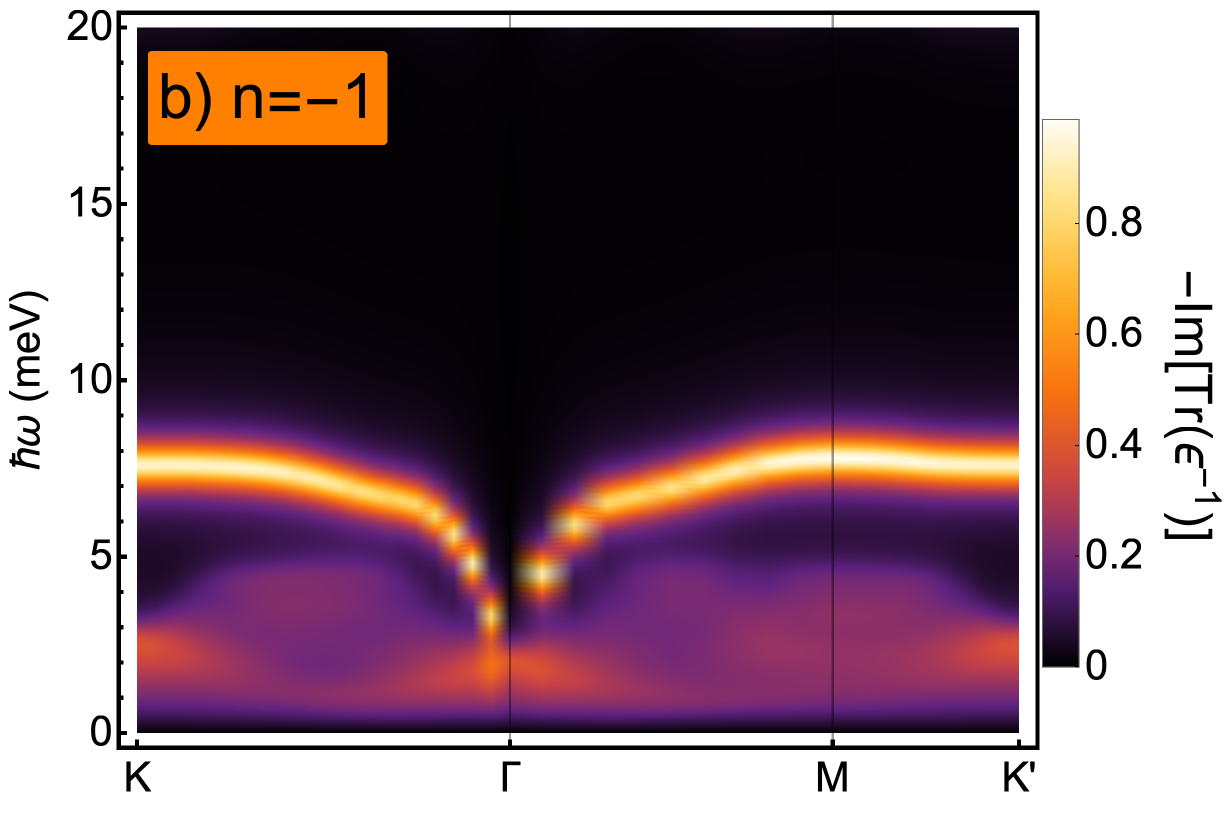}\\
\includegraphics[width=3.in]{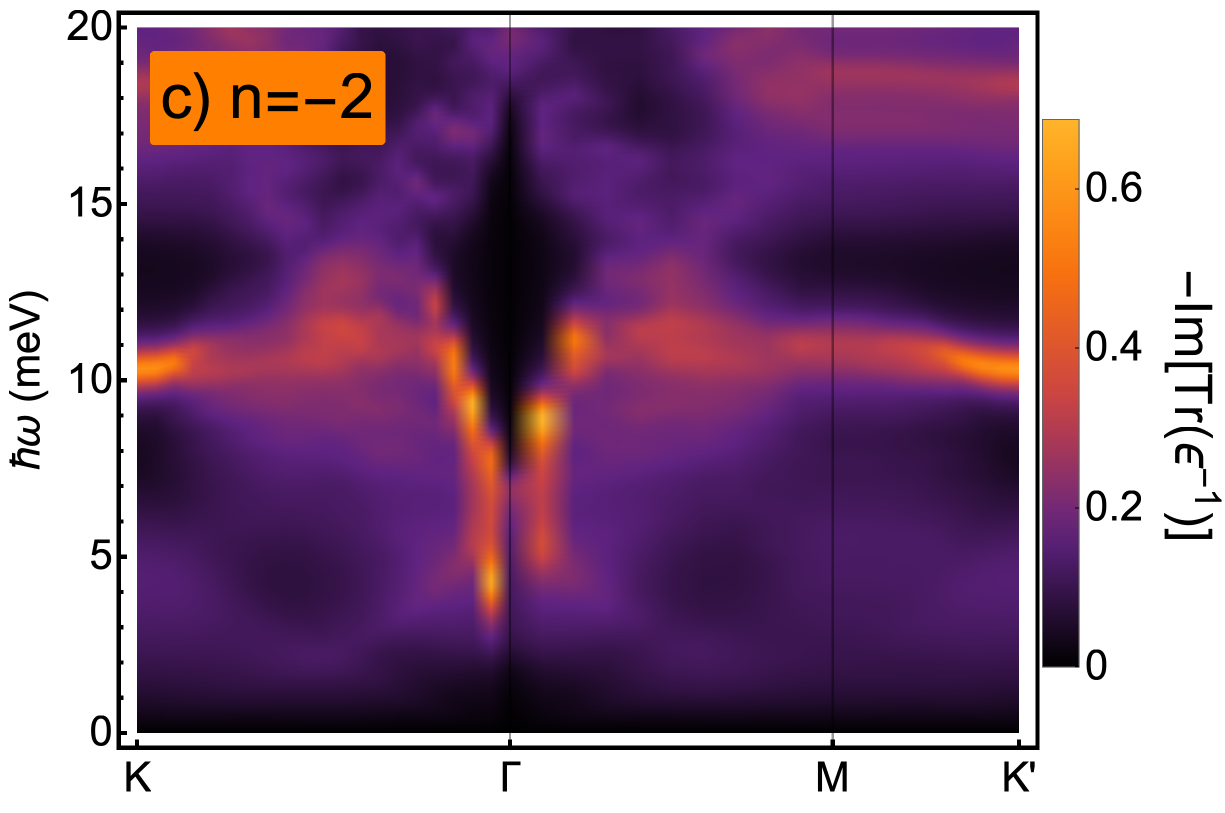}
\caption{
As in the Fig. \ref{fig:plasmon_spectrum}, but obtained by neglecting the electron-phonon coupling: $D=0$eV.
}
\label{fig:plasmon_spectrum_D=0eV}
\end{figure}
\\
\section{Electron-phonon coupling and phonon renormalization}
\label{app:phonon}
Longitudinal acoustic phonons couple to electrons through the local compression and expansion that they induce. This coupling is described by the deformation potential\cite{SA02,NGPNG09,VKG10}, $D$. We use $D = 20$ eV. Transverse acoustic phonons couple to electrons through the deformation in the interatomic distances, which leads to an effective gauge potential\cite{SA02,NGPNG09,VKG10}. This coupling is defined by the dimensionless parameter $\beta = ( a / t ) \partial t / \partial a \approx 3$, where $a$ is the interatomic distance, and $t$ is the hopping between nearest orbitals. Longitudinal phonons couple to electronic charge excitations, while transverse phonons couple to the electronic velocity operator\cite{OGM09}. 

We neglect the coupling between acoustic phonons in the two layers. This approximation is not valid when the atomic positions in the layers is significantly modified by relaxation effects, which takes place when the twist angle is small\cite{MS19,O19}. An electron state, usually delocalized through the two layers, can couple to the even and the odd combination of phonons in each layer. For longitudinal phonons, the coupling to the even mode is described by the sum of the charge fluctuations in the two layers, while the coupling to the odd mode is described by the difference. We consider only the even mode, whose coupling to electrons is significanlty larger\cite{IFGL20}. The opposite happens for the transverse phonon, as the sum of the electronic  velocities in the two layers is suppressed near a magic angle. 

We can get an estimate of the effect of the electron-phonon coupling on the phonons by considering the susceptibility in Eq. \pref{susc}. At low momenta, $| \vec{q} | \ll L^{-1}$, we take the element $\vec{G} = \vec{G}' = 0$. The energy of the renormalized phonons, $\hbar \tilde{\omega}_{\vec{q}}$ is given by:
\begin{align}
   \hbar^2 \tilde{\omega}_{\vec{q}}^2 &\approx \hbar^2 \omega_{\vec{q}}^2 \left[ 1 +\frac{\frac{D^2 }{\lambda + 2 \mu}  \chi_{0,0} ( \vec{q} , \omega_{\vec{q}} )}{1 -{\cal V}_C(\vec{q}) \chi_{0,0} ( \vec{q} , \omega_{\vec{q}} )} \right].
\end{align}
For $| \vec{q} | \rightarrow 0$ the Coulomb potential diverges: ${\cal V}_C(\vec{q})\sim 2 \pi e^2  / ( \epsilon | \vec{q} |) \rightarrow \infty$. Then, the deformation potential is screened, and phonons are weakly renormalized. For momenta such that ${\cal V}_C(\vec{q}) \chi_{0,0} ( \vec{q} , \omega_{\vec{q}} ) \ll 1$ the sound velocity, $c$, is renormalized:
\begin{align}
\tilde{c}^2 &= c^2 \left[ 1 - \frac{D^2 \left| \chi_{0,0} ( \vec{q} , \omega_{\vec{q}} ) \right|}{\lambda + 2 \mu } \right]
\end{align}
 Assuming $\chi_{0,0} ( \vec{q} , \omega_{\vec{q}} ) \approx - {\cal D} ( \epsilon_F ) \approx - {\cal N} / ( W L^2 )$, where ${\cal N} = 4$ stands for the spin and valley degeneracy,  ${\cal D} ( \epsilon_F )$  is the DOS at the Fermi level, and $W$ is an energy scale of the order of the bandwidth, we obtain a significant renormalization of the velocity of sound: $\tilde{c}^2 \approx 0.6 \times c^2$.
    \\
\section{The Kohn-Luttinger formalism for TBG}\label{APP_KL_formalism}

Because of the lack of translational invariance, it is convenient to write the linearized equation for the order parameter (OP), $\Delta$, in real space:
\begin{widetext}
\bea\label{BCS_vertex_APP}
\Delta^{i_1i_2}_{\alpha\beta}(\vec{r}_1,\vec{r}_2)=
-{\cal V}^{scr}(\vec{r}_1,\vec{r}_2)\int_\Omega\,d^2\vec{r}_3d^2\vec{r}_4
\sum_{i_3 i_4}K_BT\sum_{\omega}
\mathcal{G}^{i_1 i_3}_{\vec{r}_1 \vec{r}_3,\alpha}\left(i\hbar\omega\right)
\mathcal{G}^{ i_2 i_4}_{\vec{r}_2 \vec{r}_4,\beta}\left(-i\hbar\omega\right)
\Delta^{i_3 i_4}_{\alpha\beta}(\vec{r}_3 ,\vec{r}_4),
\eea
\end{widetext}
where
$\alpha\ne\beta$ is the index of flavor,
which encodes both the valley and spin degrees of freedom:
$\alpha=(\xi,\sigma)$,
${\cal V}$ is the screened Coulomb potential,
accounting for the RPA corrections in both the particle-hole and the electron-phonon channels,
as detailed in the App. \ref{App_Vscr} below,
$T$ is the temperature,
$\omega$ are fermionic Matsubara frequencies and
$\mathcal{G}$ is the electronic Green's function calculated in the normal state.
The Eq. \pref{BCS_vertex_APP} is represented diagrammatically in the Fig. \ref{BCS_vertex_fig},
where the wavy line denotes the screened potential and
the straight lines are the Green's functions.
\begin{figure}[h!]
\centering
\includegraphics[width=3.in]{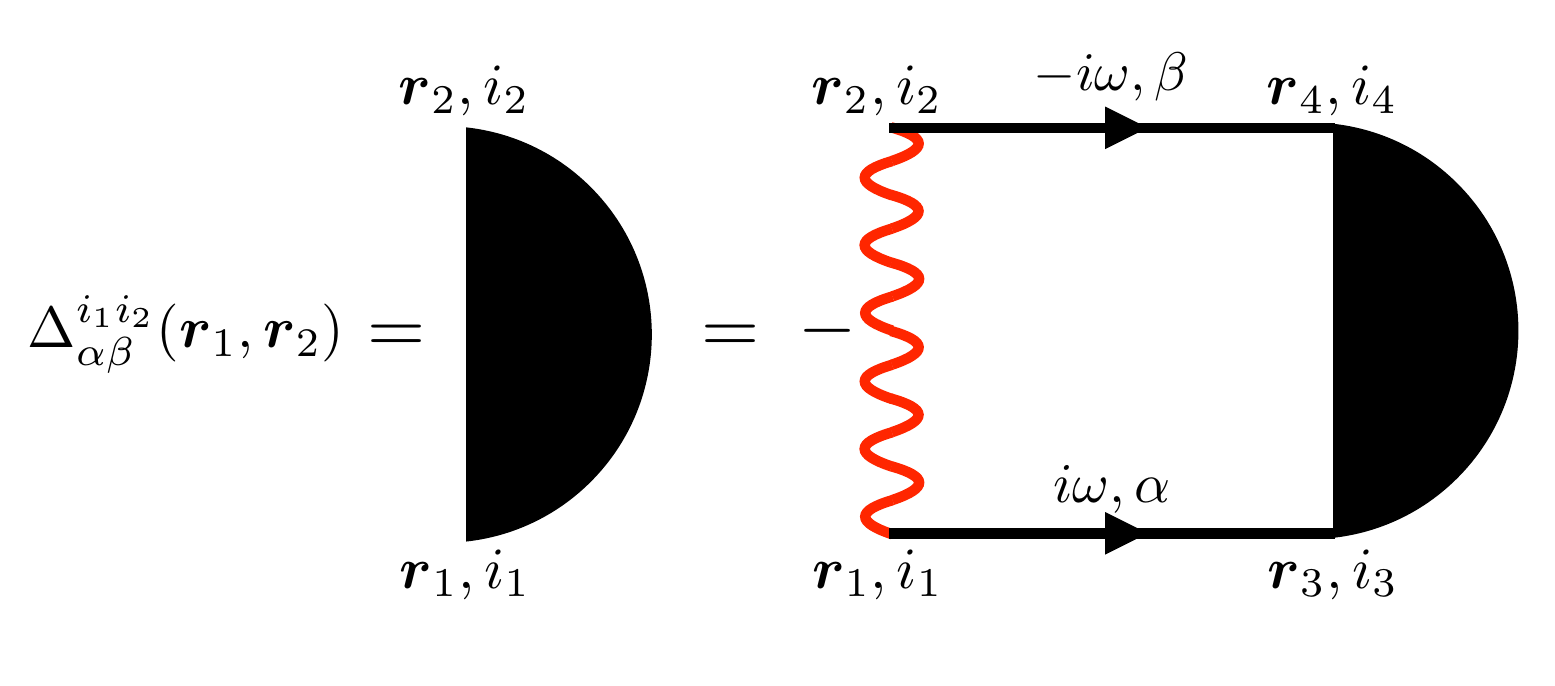}
\caption{
Diagrammatic representation of the equation \pref{BCS_vertex_APP}.
The wavy line denotes the screened potential and
the straight lines are the Green's functions.
}
\label{BCS_vertex_fig}
\end{figure}

The Green's functions can be generally written as:
\bea
\mathcal{G}^{ij}_{\vec{r}\vec{r}',\alpha}(i\hbar\omega)=\sum_{n\vec{k}}\frac{\Phi^i_{n\vec{k},\alpha}(\vec{r})\Phi^{j,*}_{n\vec{k},\alpha}(\vec{r}')}{i\hbar\omega+\mu-E_{n\vec{k},\alpha}},
\eea
where $\mu$ is the chemical potential.
Note that the Green's function does not depend explicitly on the spin index,
as the continuum Hamiltonian, Eq. \pref{HTBG}, does not.

By performing the sum over the Matsubara frequencies in the Eq. \pref{BCS_vertex_APP}, one obtains:
\begin{widetext}
\bea\label{BCS_vertex2}
\Delta^{i_1i_2}_{\alpha\beta}(\vec{r}_1,\vec{r}_2)&=&
-{\cal V}^{scr}(\vec{r}_1,\vec{r}_2)
\sum_{n_1\vec{k}_1 n_2\vec{k}_2}
\Phi^{i_1}_{n_1\vec{k}_1,\alpha}(\vec{r}_1)
\Phi^{i_2}_{n_2\vec{k}_2,\beta}(\vec{r}_2)
\left[
\frac{
f\left(-E_{n_2\vec{k}_2,\beta}+\mu\right)-f\left(E_{n_1\vec{k}_1,\alpha}-\mu\right)
}{E_{n_1\vec{k}_1,\alpha}+E_{n_2\vec{k}_2,\beta}-2\mu}\right]
\times\\
&\times&
\int_\Omega\,d^2\vec{r}_3d^2\vec{r}_4
\sum_{i_3 i_4}
\Phi^{i_3,*}_{n_1\vec{k}_1,\alpha}(\vec{r}_3)
\Phi^{i_4,*}_{n_2\vec{k}_2,\beta}(\vec{r}_4)
\Delta^{i_3i_4}_{\alpha\beta}(\vec{r}_3,\vec{r}_4),\nn
\eea
\end{widetext}
where: $f(\xi)\equiv \left(1+e^{\xi/K_BT}\right)^{-1}$ is the Fermi distribution.
Although the translational invariance is not preserved in general,
the screened potential, ${\cal V}^{scr}$, is still translationally invariant at the moir\'e scale,
which means:
${\cal V}^{scr}\left(\vec{r}_1,\vec{r}_2\right)={\cal V}^{scr}\left(\vec{r}_1+\vec{R},\vec{r}_2+\vec{R}\right)$
for any Bravais vector, $\vec{R}$, of the moir\'e lattice.
As a consequence, ${\cal V}^{scr}$ can be generally expressed in the Fourier basis as:
\begin{widetext}
\bea
{\cal V}^{scr}\left(\vec{r}_1,\vec{r}_2\right)=\frac{1}{\Omega}\sum_{\vec{q}\vec{G}_1\vec{G}_2}
{\cal V}^{scr}_{\vec{G}_1,\vec{G}_2}\left(\vec{q}\right)
e^{i\vec{q}\cdot(\vec{r}_1-\vec{r}_2)}e^{i\vec{G}_1\cdot\vec{r}_1}e^{-i\vec{G}_2\cdot\vec{r}_2},
\eea
\end{widetext}
where $\vec{q}$ belongs to the moir\'e BZ and:
${\cal V}^{scr}_{\vec{G}_1,\vec{G}_2}\left(\vec{q}+\vec{G}\right)={\cal V}^{scr}_{\vec{G}_1+\vec{G},\vec{G}_2+\vec{G}}\left(\vec{q}\right)$.
Note that the off-diagonal elements, $\vec{G}_1\ne\vec{G}_2$, are triggered by the Umklapp processes,
whereas ${\cal V}^{scr}_{\vec{G}_1,\vec{G}_2}\left(\vec{q}\right)\sim\delta_{\vec{G}_1,\vec{G}_2}$ in the translational invariant limit.
Without loss of generality, we can then assume a similar Fourier expansion for the OP:
\begin{widetext}
\bea
\Delta^{i_3i_4}_{\alpha\beta}\left(\vec{r}_3,\vec{r}_4\right)=
\frac{1}{\Omega}\sum_{\vec{q}\vec{G}_3\vec{G}_4}
\Delta^{i_3i_4}_{\alpha\beta;\vec{G}_3,\vec{G}_4}\left(\vec{q}\right)
e^{i\vec{q}\cdot(\vec{r}_3-\vec{r}_4)}e^{i\vec{G}_3\cdot\vec{r}_3}e^{-i\vec{G}_4\cdot\vec{r}_4},
\eea
\end{widetext}
which implies that only the terms with $\vec{k}_2=-\vec{k}_1$ survive in the rhs of the Eq. \pref{BCS_vertex2}.
Defining:
\begin{subequations}
\bea
\Delta^{n_1n_2}_{\alpha\beta}(\vec{q})&\equiv&\\
\int_\Omega\,d^2\vec{r}_3d^2\vec{r}_4
\sum_{i_3 i_4}&&
\Phi^{i_3,*}_{n_1\vec{q},\alpha}(\vec{r}_3)
\Phi^{i_4,*}_{n_2-\vec{q},\beta}(\vec{r}_4)
\Delta^{i_3i_4}_{\alpha\beta}(\vec{r}_3,\vec{r}_4),\nn\\
\tilde{\Delta}^{n_1n_2}_{\alpha\beta}(\vec{q})&\equiv&
\Delta^{n_1n_2}_{\alpha\beta}(\vec{q})\times\\
&\times&
\sqrt{
\frac{
f\left(-E_{n_2-\vec{q},\beta}+\mu\right)-f\left(E_{n_1\vec{q},\alpha}-\mu\right)
}{E_{n_2-\vec{q},\beta}+E_{n_1\vec{q},\alpha}-2\mu}
},\nn
\eea
\end{subequations}
we project the Eq. \pref{BCS_vertex2} on the Bloch's eigenstates,
which gives the following equation for $\tilde{\Delta}^{n_1n_2}_{\alpha\beta}(\vec{q})$:
\bea\label{BCS_vertex_qspace}
\tilde{\Delta}^{m_1m_2}_{\alpha\beta}(\vec{k})=
\sum_{n_1n_2}\sum_{\vec{q}}\Gamma^{m_1m_2}_{n_1n_2;\alpha\beta}(\vec{k},\vec{q})
\tilde{\Delta}^{n_1n_2}_{\alpha\beta}(\vec{q}),
\eea
with:
\begin{widetext}
\bea\label{kernel}
\Gamma^{m_1m_2}_{n_1n_2;\alpha\beta}(\vec{k},\vec{q})&=&
-\frac{1}{\Omega}\sum_{\vec{G_1}\vec{G_1}'}\sum_{\vec{G_2}\vec{G_2}'}\sum_{i_1i_2}
{\cal V}^{scr}_{\vec{G}_1-\vec{G}_1',\vec{G}_2-\vec{G}_2'}\left(\vec{k}-\vec{q}\right)
\phi^{i_1,*}_{m_1\vec{k},\alpha}(\vec{G}_1)
\phi^{i_2,*}_{m_2-\vec{k},\beta}\left(\vec{G}_2'\right)
\phi^{i_1}_{n_1\vec{q},\alpha}\left(\vec{G}_1'\right)
\phi^{i_2}_{n_2-\vec{q},\beta}(\vec{G}_2)\times\nn\\
&\times&
\sqrt{
\frac{
f\left(-E_{m_2-\vec{k},\beta}+\mu\right)-f\left(E_{m_1\vec{k},\alpha}-\mu\right)
}{E_{m_2-\vec{k},\beta}+E_{m_1\vec{k},\alpha}-2\mu}}\times
\sqrt{
\frac{
f\left(-E_{n_2-\vec{q},\beta}+\mu\right)-f\left(E_{n_1\vec{q},\alpha}-\mu\right)
}{E_{n_2-\vec{q},\beta}+E_{n_1\vec{q},\alpha}-2\mu}}.
\eea
\end{widetext}
The condition for the onset of superconductivity is that the kernel $\Gamma$ has the eigenvalue 1.
This defines the critical temperature, $T_c$, as the one at which the largest eigenvalue of $\Gamma$ is equal to 1.

Because the eigenfunctions do not depend explicitly on the spin index,
we can identify two different kinds of OP,
depending if $\alpha$ and $\beta$ share the same or opposite valley indices.
These two OP's describe intra-valley or inter-valley superconductivity, respectively. In the present work we focus on the inter-valley case, as we checked that the intra-valley superconductivity is less robust.

In order to diagonalize the kernel $\Gamma$, we project the Eq. \pref{BCS_vertex_qspace} onto the two bands in the middle of the spectrum, that give the main contribution to superconductivity.
For each filling, we include into $\Gamma$ the Hartree corrections on top of the non-interacting band structure,
as detailed in the Refs.~\cite{Guinea_pnas18,cea_prb19}.

\section{Computation of the screened potential}\label{App_Vscr}
Here we provide the details concerning the calculation of the screened potential in the reciprocal space:
${\cal V}^{scr}_{\vec{G}_1,\vec{G}_2}\left(\vec{q}\right)$.
First, we introduce the unscreened Coulomb potential:
\bea
{\cal V}_C(\vec{q})=2\pi e^2\frac{\tanh\left(d_g|\vec{q}|\right)}{\epsilon|\vec{q}|},
\eea
where $e$ is the electron charge, $d_g$ is the distance of the sample from a metallic gate and $\epsilon$ is the relative dielectric constant of the environment.
In this work we use: $\epsilon=10$ and $d_g=40$nm.
We consider the effect of the strain in TBG by means of longitudinal acoustic phonons coupling to the charge density via the deformation potential:
$v_d(\vec{r})=-D s(\vec{r})$, where $s(\vec{r})$ is the local strain and $D$ is the electron-phonon coupling, for which we use: $D=20$eV.
We compute the screened potential, including the RPA corrections induced by both the particle-hole excitations and the electron-phonon coupling.
In matrix notation, the expression for the inverse of ${\cal V}^{scr}$ is given by:
\bea\label{inverseVscr}
\left[{\cal V}^{scr}(\vec{q})\right]^{-1}&=&\left[\hat{{\cal V}}_C(\vec{q})\right]^{-1}-\chi^0(\vec{q})+\\
&+&g\chi^0(\vec{q})\left[\mathbb{1}+g\chi^0(\vec{q})\right]^{-1}\chi^0(\vec{q}),\nn
\eea
where the entries are indexed by the reciprocal lattice vectors,
$\left[\hat{{\cal V}}_C(\vec{q})\right]_{\vec{G}_1,\vec{G}_2}\equiv {\cal V}_C\left(\vec{q}+\vec{G}_1\right)\delta_{\vec{G}_1,\vec{G}_2}$,
$g=\frac{D^2}{\lambda+2\mu}$,
 with $\lambda=3.25$eV\AA$^{-2}$, $\mu=9.44$eV\AA$^{-2}$ the Lam\'e coefficients of monolayer graphene,
and $\chi^0(\vec{q})$ is the static density-density response function,
which is given by:
\begin{widetext}
\bea
\chi^0_{\vec{G}_1,\vec{G}_2}\left(\vec{q}\right)&=&\frac{1}{\Omega}\sum_{\vec{k}\vec{G}'_1\vec{G}'_2}\sum_{n_1n_2}\sum_{ij\alpha}
\phi^i_{n_1\vec{k}+\vec{q},\alpha}\left(\vec{G}'_1+\vec{G}_1\right)\phi^{i,*}_{n_2\vec{k},\alpha}\left(\vec{G}'_1\right)
\phi^{j,*}_{n_1\vec{k}+\vec{q},\alpha}\left(\vec{G}'_2+\vec{G}_2\right)\phi^j_{n_2\vec{k},\alpha}\left(\vec{G}'_2\right)\times\\
&\times&
\frac{f\left( E_{n_2\vec{k},\alpha}-\mu \right)-f\left( E_{n_1\vec{k}+\vec{q},\alpha}-\mu \right)}
{ E_{n_2\vec{k},\alpha}-E_{n_1\vec{k}+\vec{q},\alpha}}.
\nn
\eea
\end{widetext}
To compute $\chi^0$ numerically, we include the 10 bands closest to the charge neutrality point.

The Fig. \ref{Vscr} shows the screened potential in real space, obtained upon Fourier transforming the inverse of the Eq. \pref{inverseVscr},
at the magic angle: $\theta=1.085^\circ$, and at half filling.
The red (blue) line displays the effective interaction between two electrons close to an AA-stacked (BA-stacked) region of the moir\'e unit cell,
as a function of the inter-particle distance, $x$, according to:
${\cal V}^{scr}_{AA/AB}(x)\equiv {\cal V}^{scr}\left(\vec{R}_{AA/AB}+x,\vec{R}_{AA/AB}\right)$.
Note the attractive behavior close to: $x\simeq2$nm.
\begin{figure}[h!]
\centering
\includegraphics[width=3.in]{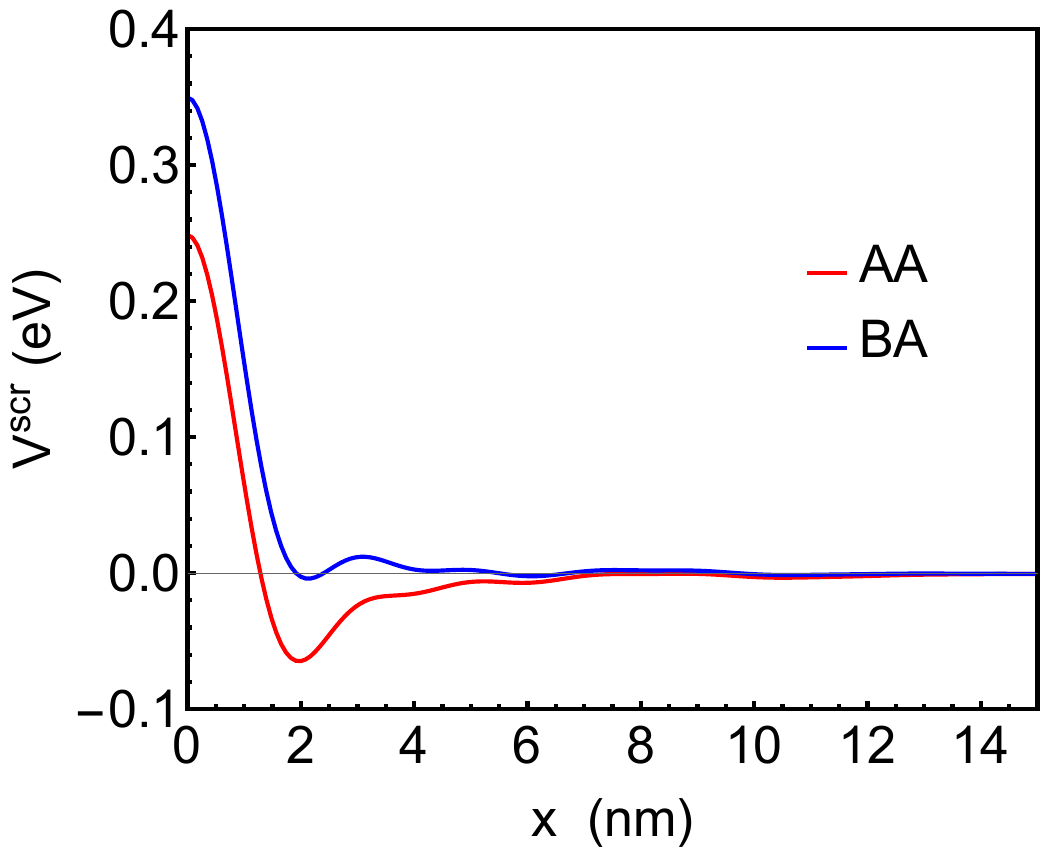}
\caption{
Screened potential in real space, obtained for $\theta=1.085^\circ$ at half filling,
in the neighboring of an AA- and a BA-stacked region of the moir\'e unit cell.
}
\label{Vscr}
\end{figure}

\section{Analytical approximation to the screened potential}\label{App_anal_Vscr}

For simplicity, we consider here only the screening of the Coulomb potential, ${\cal V}_C ( \vec{q} )$. We consider the two central bands of TBG. The inclusion of the electron-phonon interaction can be carried out along the same lines, although it will complicate the expressions.

The electronic polarizability is a matrix, whose entries are reciprocal lattice vectors:
\begin{align}
\chi_{\vec{G} , \vec{G}'} ( \vec{q} , \omega )&=\frac{1}{\Omega} \sum_{\vec{k}} \cal{M}^*_{\vec{G},\vec{k},\vec{k}+\vec{q}} \cal{M}_{\vec{G}',\vec{k},\vec{k}+\vec{q}} \frac{ {\it n}_{\vec{k}}- {\it n}_{\vec{k} + \vec{q}} }{\omega - \epsilon_{\vec{k}+\vec{q}} + \epsilon_{\vec{k}}}
\end{align}
where $n_{\vec{k}}$ stands for the Fermi-Dirac distribution, $\epsilon_{\vec{k}}$ is the energy of a state with momentum $\vec{k}$, and where, for simplicity, we have omitted that the summation runs over the two central bands. Finally, there is an overlap factor:
\begin{align}
{\cal M}_{\vec{G} , \vec{k} , \vec{k} + \vec{q}} &= \int_{A_C} d^2 \vec{r} e^{i \vec{G} \vec{r}} u^*_{\vec{k} + \vec{q}} ( \vec{r} ) u_{\vec{k}} ( \vec{r} )
\label{factor0}
\end{align}
where $A_C$ is the unit cell in real space, and $u_{\vec{k}} ( \vec{r} )$ is the periodic part of the wavefunction, $\psi_{\vec{k}} ( \vec{r} ) = u_{\vec{k}} ( \vec{r} ) e^{i \vec{k} \vec{r}}$.

The dielectric function is the matrix:
\begin{align}
\epsilon_{\vec{G} , \vec{G}'} ( \vec{q} , \omega ) &= \delta_{\vec{G} , \vec{G}'} - {\cal V}_C\left(\vec{G}+\vec{q}\right) \chi_{\vec{G} , \vec{G}'} ( \vec{q} , \omega )
\end{align}
The inversion of this matrix, $\epsilon^{-1}_{\vec{G} , \vec{G}'} ( \vec{q} , \omega )$ is greatly simplified if we make the assumption that the overlap factor can be factorized:
\begin{align}
{\cal M}_{\vec{G} , \vec{k} , \vec{k} + \vec{q}} &= f_{\vec{G}} \times g ( \vec{k} , \vec{k} + \vec{q} )
\label{factor}
\end{align}
This approximation can be justified using Wannier functions, and assuming that the operator $e^{i \vec{G} \vec{r}}$ is diagonal in the Wannier basis\cite{IFGL20}. It is equivalent to the "flat metric condition" used in\cite{Betal20b}. Note that this approximation implies that a periodic potential, such as the Hartree potential, affects equally all states in the BZ, so that it cannot describe the changes in the shape of the bands in the Hartree approximation\cite{Guinea_pnas18}. We discuss on how to include these effects at the end of this section.

The normalization of the wavefunctions implies that $f_{\vec{G} = 0} = g ( \vec{k} , \vec{k} ) = 1$.
Using Eq. \pref{factor}, we obtain:
\begin{align}
\chi_{\vec{G} , \vec{G}'} ( \vec{q} , \omega )&= f_{\vec{G}} f_{\vec{G}'} \tilde{\chi} ( \vec{q} , \omega )
\label{chitilde}
\end{align}
where we have included the factors $g ( \vec{k} , \vec{q} )$ into the definition of $\tilde{\chi}$:
\begin{align}
\tilde{\chi} ( \vec{q} , \omega )&= \sum_{\vec{k}} \left| g ( \vec{k} , \vec{k} + \vec{q} ) \right|^2  \frac{{\it n}_{\vec{k}}-{\it n}_{\vec{k} + \vec{q}}}{\omega+ \epsilon_{\vec{k}} - \epsilon_{\vec{k}+\vec{q}} }
\end{align}
The dielectric function becomes:
\begin{align}
\epsilon_{\vec{G} , \vec{G}'} ( \vec{q} , \omega ) &= \delta_{\vec{G} , \vec{G}'} - {\cal V}_C\left(\vec{G}+\vec{q}\right) f_{\vec{G}} f_{\vec{G}'} \tilde{\chi}_{\vec{G} , \vec{G}'} ( \vec{q} , \omega )
\end{align}
The elements $\epsilon^{-1}_{\vec{G} , \vec{G'}} ( \vec{q} , \omega )$ satisfy:
\begin{align}
\epsilon^{-1}_{\vec{G} , \vec{G'}} ( \vec{q} , \omega ) &-&\nn\\
-{\cal V}_C\left(\vec{G} + \vec{q}\right) f_{\vec{G}}  \tilde{\chi} ( \vec{q} \omega ) \sum_{\vec{G}''} f_{\vec{G}''} \epsilon^{-1}_{\vec{G}'' , \vec{G'}} ( \vec{q} , \omega ) &=& \delta_{\vec{G} , \vec{G}'}
\label{epsiloninv}
\end{align}
We define:
\begin{align}
F_{\vec{G}} ( \vec{q} , \omega ) &= \sum_{\vec{G}'} f_{\vec{G}'} \epsilon^{-1}_{\vec{G}' , \vec{G}} ( \vec{q} , \omega )
\end{align}
Then:
\begin{align}
F_{\vec{G}'} - \tilde{\chi} ( \vec{q} , \omega )  \sum_{\vec{G}''} {\cal V}_C\left(\vec{G}'' + \vec{q}\right) f_{\vec{G}''}^2 F_{\vec{G}'} &= f_{\vec{G}'}
\end{align}
and
\begin{align}
F_{\vec{G}'} = \frac{f_{\vec{G}'}}{1 -  \tilde{\chi} ( \vec{q} , \omega ) \sum_{\vec{G}''} {\cal V}_C\left(\vec{G}'' + \vec{q}\right) f_{\vec{G}''}^2}
\end{align}
Inserting this expression in Eq. \pref{epsiloninv} we find:
\begin{align}
\epsilon^{-1}_{\vec{G} , \vec{G}'} ( \vec{q} , \omega ) &= \delta_{\vec{G} , \vec{G}'} + \frac{{\cal V}_C\left(\vec{G} + \vec{q}\right) f_{\vec{G}} f_{\vec{G}'} \tilde{\chi} ( \vec{q} , \omega ) }{1 -  \tilde{\chi} ( \vec{q} , \omega ) \sum_{\vec{G}''} {\cal V}_C\left(\vec{G}'' + \vec{q}\right) f_{\vec{G}''}^2}
\label{epsiloninvp}
\end{align}
The screened potential is:
\begin{widetext}
\begin{align}
{\cal V}^{scr}_{\vec{G} , \vec{G}'} ( \vec{q} , \omega ) = \epsilon^{-1}_{\vec{G} , \vec{G}'} ( \vec{q} , \omega ) {\cal V}_C\left(\vec{G}' + \vec{q}\right) = 
{\cal V}_C\left(\vec{G} + \vec{q}\right) \delta_{\vec{G} , \vec{G}'} + \frac{{\cal V}_C\left(\vec{G} + \vec{q}\right) {\cal V}_C\left(\vec{G}' + \vec{q}\right) f_{\vec{G}} f_{\vec{G}'} \tilde{\chi} ( \vec{q} , \omega ) }{1 -  \tilde{\chi} ( \vec{q} , \omega ) \sum_{\vec{G}''} {\cal V}_C\left(\vec{G}'' + \vec{q}\right) f_{\vec{G}''}^2}
\label{vscr2}
\end{align}
\end{widetext}
The pairing diagram which describes the scattering of a Cooper pair with momenta $\{ \vec{k} , - \vec{k} \}$ to a Cooper pair with momenta  $\{ \vec{k} + \vec{q} , - \vec{k} - \vec{q} \}$ involves extra factors $f_{\vec{G}} g ( \vec{k} , \vec{k} + \vec{q} )$:
\begin{align}
\Gamma_{\vec{k} , \vec{k} + \vec{q}} ( \omega ) &=- g^2 ( \vec{k} , \vec{k} + \vec{ q} )\sum_{\vec{G} , \vec{G}'} f_{\vec{G}} f_{\vec{G}'} {\cal V}^{scr} _{\vec{G} , \vec{G}'} ( \vec{q} , \omega ) = \nonumber \\
&=-  g^2 ( \vec{k} , \vec{k} + \vec{ q} ) \frac{\sum_{\vec{G}} {\cal V}_C ( \vec{G} + \vec{q} ) f_{\vec{G}}^2}{1 - \tilde{\chi} ( \vec{q} , \omega ) \sum_{\vec{G}} {\cal V}_C ( \vec{G} + \vec{q} ) f_{\vec{G}}^2} 
\label{kernel2}
\end{align}
This expression is equivalent to the definition of an effective potential which describes the Umklapp processes, ${\cal V}_{eff} ( \vec{q} ) = \sum_{\vec{G}} {\cal V}_C ( \vec{G} + \vec{q} ) f_{\vec{G}}^2$. This kernel is negative, i. e. repulsive, for $\omega = 0)$ and all values of $\vec{k}$ and $\vec{k} + \vec{q}$. It does not favor superconductivity, in agreement with the results in\cite{Betal20b}.

The factorization in Eq. \pref{factor} implies the replacement of expectation values of the type $\langle \vec{k} |  e^{i \vec{G} \vec{r}} | \vec{k} \rangle$ by their average over the BZ, or, alternatively, this approximation implies the neglect of the expectation value of the operator $e^{i \vec{G} \vec{r}}$ between Wannier functions centered at different sites.The approximation fails for values of $\vec{G}$ in the first star of reciprocal lattice vectors\cite{Betal20b}. Eq. \pref{chitilde} needs to be replaced by:
\begin{align}
\chi_{\vec{G} , \vec{G}'} ( \vec{q} , \omega )&= f_{\vec{G}} f_{\vec{G}'} \tilde{\chi} ( \vec{q} , \omega ) + \delta \chi_{\vec{G} , \vec{G}'} ( \vec{q} , \omega )
\label{chitildep}
\end{align}
The correction $\delta \chi_{\vec{G} , \vec{G}'} ( \vec{q} , \omega )$ is not negligible when $\vec{G} , \vec{G}'$ lie in the first star of reciprocal lattice vectors.

We can include the effect of $\delta \chi_{\vec{G} , \vec{G}'} ( \vec{q} , \omega )$ in $\epsilon^{-1}_{\vec{G} , \vec{G}'} ( \vec{q} , \omega )$ perturbatively. To lowest order, we obtain:
\begin{widetext}
\begin{align}
\delta \epsilon^{-1}_{\vec{G} , \vec{G}'} ( \vec{q} , \omega ) &\approx \sum_{\vec{G}'' , \vec{G}'''}  \epsilon^{-1}_{\vec{G} , \vec{G}''} ( \vec{q} , \omega ) {\cal V}_C ( \vec{G}'' + \vec{q} ) \delta \chi_{\vec{G}'' , \vec{G}'''} ( \vec{q} , \omega )  \epsilon^{-1}_{\vec{G}''' , \vec{G}'} ( \vec{q} , \omega ) 
\nonumber \\
\delta {\cal V}^{scr}_{\vec{G} , \vec{G}'} ( \vec{q} , \omega )&\approx \sum_{\vec{G}'' , \vec{G}'''} {\cal V}^{scr}_{\vec{G} , \vec{G}''} ( \vec{q} , \omega ) \delta \chi_{\vec{G}'' , \vec{G}'''} ( \vec{q} , \omega ) {\cal V}^{scr}_{\vec{G}''' , \vec{G}'} ( \vec{q} , \omega )
\label{vscr3}
\end{align}
\end{widetext}
where $\epsilon^{-1}_{\vec{G} , \vec{G}'} ( \vec{q} , \omega )$ and ${\cal V}^{scr}_{\vec{G} , \vec{G}'} (\vec{q} , \omega )$ as defined in Eq. \pref{epsiloninvp} and Eq. \pref{vscr2}.

The correction to the screening potential in Eq. \pref{vscr3} depends on the deviations of the functions $| {\cal M}_{\vec{G} , \vec{k} , \vec{k} + \vec{q}} |^2$ in Eq. \pref{factor0} from their average over the BZ, Eq. \pref{factor}. For the first magic angle, these corrections are largest for the first star of reciprocal lattice vectors, $| \vec{G} | = | \vec{G}' | = (4 \pi ) / ( \sqrt{3} L )$. For $| \vec{q} | \simeq | \vec{G} |$ these corrections are largest and positive for $\vec{k}$ in the vicinity of the edges of the BZ, and largest and negative when $\vec{k}$ is near the $\Gamma$ point at the center of the BZ. Finally, the weight of $| {\cal M}_{\vec{G} , \vec{k} , \vec{k} + \vec{q}} |^2$
 on the value of $\delta \chi$ will be highest when $\vec{k}$ is near the Fermi surface. The two factors other than $\delta{\chi}$ in Eq. \pref{vscr3} are of the order of the square of the screened potential. Hence, the correction to the potential in Eq. \pref{vscr3} is attractive (negative) when the Fermi surface is near the edge of the BZ, and repulsive (positive) when the Fermi surface is near $\Gamma$, the center of the BZ.
 
 We can further estimate the validity of the approximation in Eq. \pref{factor} by comparing numerical results with predictions obtained using this equation for averages: 
 \begin{align}
{\cal F}_{\vec{G} , \vec{G}'}  ( \vec{q} )&= \left. \overline{{\cal M} ^*_{\vec{G}}  ( \vec{k} , \vec{k} + \vec{q} ) {\cal M}_{\vec{G}'} ( \vec{k} , \vec{k} + \vec{q} )} \right|_{\small  \vec{k} \in BZ  }
\label{average}
\end{align}
which enter in the response function and in the kernel $\Gamma ( \vec{k} , \vec{k} + \vec{q} )$.

The approximation in Eq. \pref{factor} implies that:
\begin{align}
{\cal F}_{\vec{G} , \vec{G}'}  ( \vec{q} ) &\approx f_{\vec{G}} f_{\vec{G}'} \left. \overline{g^2 ( \vec{k} , \vec{k} + \vec{g} ) } \right|_{\small  \vec{k} \in BZ  } 
\label{approx2}
\end{align}
We first consider ${\cal F}_{\vec{G} , \vec{G}'} ( \vec{q} = 0  )$.  The approximation in Eq. \pref{factor} gives ${\cal F}_{\vec{G} , \vec{G}'} ( \vec{q} = 0  ) = f_{\vec{G}} f_{\vec{G}'}$.

The form factors in Eq. \pref{average} with $\vec{G} \ne \vec{G}'$ enter in the effective interaction always associated to polarization bubbles, so that they describe an attractive contribution to the total interaction. Hence, the approximation in Eq. \pref{factor} underestimates the attractive interaction between quasiparticles.
Nomerical results for  ${\cal F}_{\vec{G} , \vec{G}'} ( \vec{q} = 0  )$ calculated at a magic angle are shown in Table[\ref{table_q0}] (the reciprocal lattice vectors are parametrized as $\vec{G}_{m,n} = (4 \pi ) / ( \sqrt{3} L ) [ m ( \sqrt{3} / 2 , 1/2 ) + n ( \sqrt{3} / 2 , -1/2 ) ] $ ) and we use the definition $| \vec{G} | = \sqrt{m^2 + n^2 + mn}$.  The results have been obtained for states in the valence band of TBG at a magic angle.

Using the approximation in Eq. \pref{factor}, the entries with $\{ \vec{G} , \vec{G}' \} = \{ 0 , \vec{G}' \}$ in Table[\ref{table_q0}] lead to  $f_{1} \approx 0.86 , f_{\sqrt{3}} \approx 0.24$. These results imply that ${\cal F}_{( 1, 0 ) , ( 1, 0)} ( \vec{q} = 0 ) = f^2_{1} \approx 0.74 , \, {\cal F}_{ ( 1, 0 ) , ( 1, 1)} ( \vec{q} = 0 ) = f_{ 1} \times  f_{\sqrt{3}}\approx 0.20 ,  \, {\cal F}_{( 1, 1 ) , ( 1, 1)} ( \vec{q} = 0 ) = f^2_{\sqrt{3}} \approx 0.056$. The corresponding values extracted from Table[\ref{table_q0}] are $\, {\cal F}_{( 1, 0 ) ,  ( 1, 0)} ( \vec{q} = 0 ) = 0.78$ , ${\cal F}_{ ( 1, 0 ) , ( 1, 1)} ( \vec{q} = 0 ) =0.24$ and ${\cal F}_{( 1, 1 ) ,  ( 1, 1)} ( \vec{q} = 0 ) = 0.067$.

\begin{table}[H]
\begin{center}
\begin{tabular}{||c|c||c||}
\hline
\hline
$\vec{G} , | \vec{G} |$ & $\vec{G}' , | \vec{G}' |$ &${\cal F}_{\vec{G} , \vec{G}'} ( \vec{q} = 0 )$ \\
\hline
$( 0 , 0 ) , 0 $ & ( 0 , 0 ) , 0 & 1 \\
$( 0 , 0 ) , 0$ & ( 1 , 0 ) , 1 & 0.861 \\
$( 0 , 0 ) , 0$ & ( 1 , 1 ) , $\sqrt{3}$ & 0.238 \\
$( 1 , 0 ) , 1$ & ( 1 , 0 ) , 1& 0.785 \\
$( 1 , 0 ) , 1$ & ( 0 , 1 ) , 1 & 0.761 \\
$(1 , 0 ) , 1$ & ( -1 , 1 ) , 1 & 0.761 \\
$( 1 , 0 ) , 1$ & ( 1 , 1 ) , $\sqrt{3}$ & 0.238 \\
$( 0 , 1 ) , 1$ & ( 1 , 1 ) , $\sqrt{3}$ & 0.223 \\
$( -1 , 1 ) , 1$ & ( 1 , 1 ) , $\sqrt{3}$ & 0.223 \\
$( 1 , 1 ) , \sqrt{3} $ & ( 1 , 1 ) , $\sqrt{3}$& 0.067 \\
$( 1 , 1 )  , \sqrt{3}$ & ( -2 , 1 ) , $\sqrt{3}$ & 0.063 \\
$( 1 , 1 )  , \sqrt{3}$ & ( 1 , -2 ) , $\sqrt{3}$ & 0.063 \\
\hline
\hline
\end{tabular}
\end{center}
\caption{\label{table_q0} Numerical results for the function  ${\cal F}_{\vec{G} , \vec{G}'} ( \vec{q} = 0  )$ in Eq. \pref{average} for different values of $\vec{G}$ and $\vec{G}'$. The parameters define TBG at a magic angle. }
\end{table}

We now average over all transitions within the BZ, by defining:
\begin{align}
\tilde{{\cal F}}_{\vec{G} , \vec{G}'} &=  \left. \overline{\overline{{\cal M} ^*_{\vec{G}}  ( \vec{k} , \vec{k} + \vec{q} ) {\cal M}_{\vec{G}'} ( \vec{k} , \vec{k} + \vec{q} )}} \right|_{\scriptsize  \begin{array}{c} {\vec{k} \in BZ} \\ { \vec{k} + \vec{q} \in BZ} \end{array}  }
\label{average2}
\end{align}

Numerical results for this function are shown in Table[\ref{table_qn0}].

\begin{table}[H]
\begin{center}
\begin{tabular}{||c|c||c||}
\hline
\hline
$\vec{G} , | \vec{G} |$ & $\vec{G}' , | \vec{G}' |$ &$\tilde{{\cal F}}_{\vec{G} , \vec{G}'}$ \\
\hline
$( 0 , 0 ) , 0 $ & ( 0 , 0 ) , 0 & 0.290 \\
$( 0 , 0 ) , 0$ & ( 1 , 0 ) , 1 & 0.303 \\
$( 0 , 0 ) , 0$ & ( 1 , 1 ) , $\sqrt{3}$ & 0.100 \\
$( 1 , 0 ) , 1$ & ( 1 , 0 ) , 1& 0.354 \\
$( 1 , 0 ) , 1$ & ( 0 , 1 ) , 1 & 0.328 \\
$(1 , 0 ) , 1$ & ( -1 , 1 ) , 1 & 0.328 \\
$( 1 , 0 ) , 1$ & ( 1 , 1 ) , $\sqrt{3}$ & 0.100 \\
$( 0 , 1 ) , 1$ & ( 1 , 1 ) , $\sqrt{3}$ & 0.121 \\
$( -1 , 1 ) , 1$ & ( 1 , 1 ) , $\sqrt{3}$ & 0.121 \\
$( 1 , 1 ) , \sqrt{3} $ & ( 1 , 1 ) , $\sqrt{3}$& 0.049 \\
$( 1 , 1 )  , \sqrt{3}$ & ( -2 , 1 ) , $\sqrt{3}$ & 0.039 \\
$( 1 , 1 )  , \sqrt{3}$ & ( 1 , -2 ) , $\sqrt{3}$ & 0.039 \\
\hline
\hline
\end{tabular}
\end{center}
\caption{\label{table_qn0} Numerical results for the function  $\tilde{{\cal F}}_{\vec{G} , \vec{G}'} $ in Eq. \pref{average2} for different values of $\vec{G}$ and $\vec{G}'$. The parameters define TBG at a magic angle. }
\end{table}

The approximation in Eq. \pref{factor} gives:
\begin{align}
\tilde{{\cal F}}_{\vec{G} , \vec{G}'} &\approx f_{\vec{G}} f_{\vec{G}'} \left.  \overline{\overline{g^2 ( \vec{k} , \vec{k} + \vec{q} )}}  \right|_{\scriptsize  \begin{array}{c} {\vec{k} \in BZ} \\ { \vec{k} + \vec{q} \in BZ} \end{array}  } = \nonumber \\
& = f_{\vec{G}} f_{\vec{G}'} \tilde{{\cal F}}_{\vec{G}=0 , \vec{G}'=0}
\label{approx3}
\end{align}
This approximation deviates significantly from the results in Table[\ref{table_qn0}]. The approximation underestimates the form factor defined by states which are not close inside the BZ. 
\section{Effect of the external screening on the pairing}\label{App_screening}

Fig. \ref{screening_dependence} compares the critical temperatures obtained for different environmental screenings: $\epsilon=10$ (a) and $\epsilon=3.5$ (b), as functions of the electronic filling.
As is evident, decreasing the screening lowers $T_C$, especially at finite filling. This is mainly due to the fact that, the smaller is the screening, the stronger is the Coulomb interaction, and consequently the wider are the bands at finite filling, which dramatically suppresses the van Hove singularities.  
\begin{figure}[h!]
\centering
\includegraphics[width=3.2in]{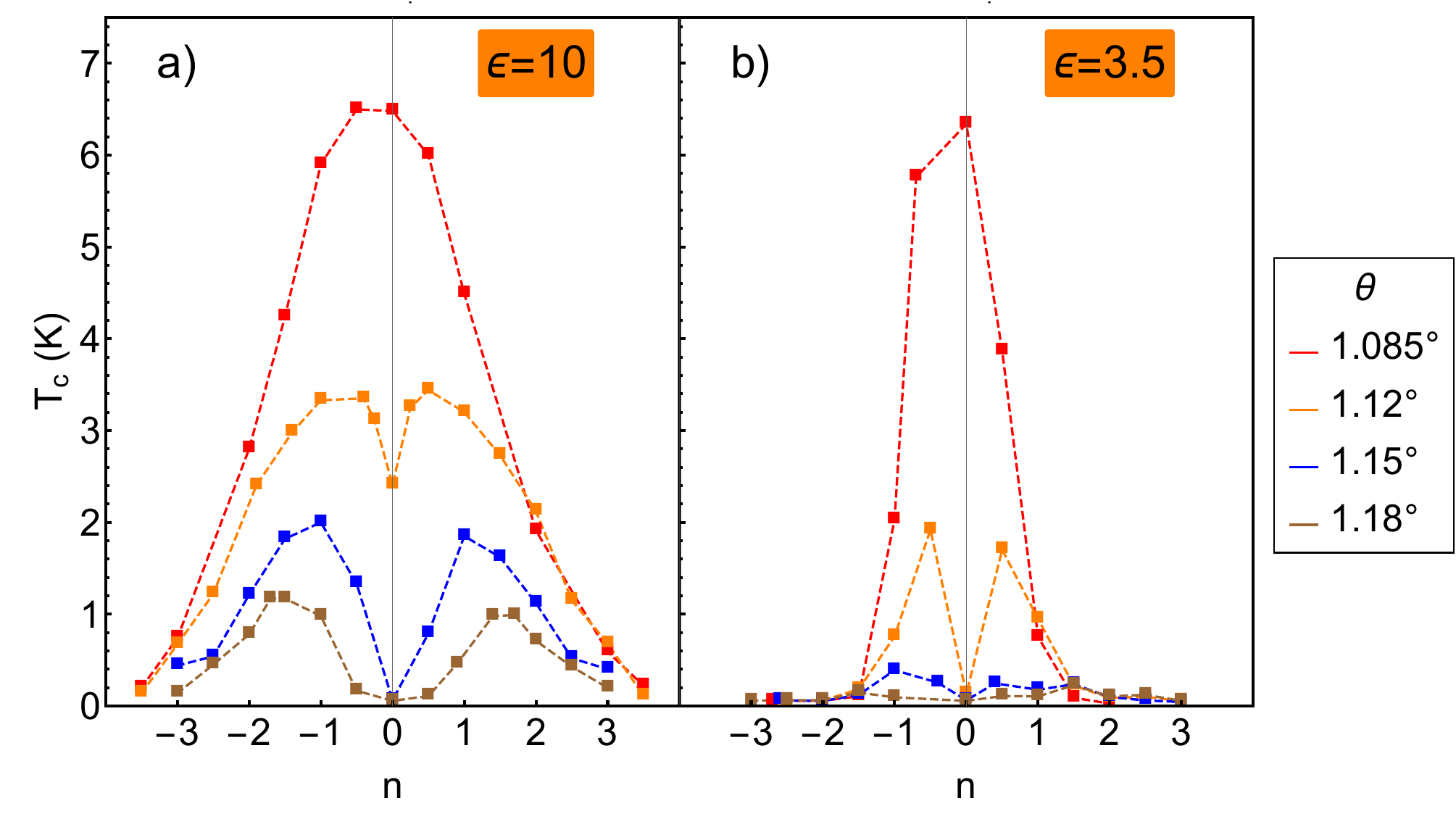}
\caption{
Critical temperatures obtained for different environmental screenings: $\epsilon=10$ (a) and $\epsilon=3.5$ (b), as functions of the electronic filling.
}
\label{screening_dependence}
\end{figure}
It's worth noting that this tendency is inverted by suppressing the electron-phonon coupling.
Fig. \ref{screening_dependence_D=0} shows the critical temperatures obtained at $\theta=1.085^\circ$ for $\epsilon=10$ (red line) and $\epsilon=3.5$ (blue line) and $D=0$. Although the values of $T_C$ are nearly two order of magnitude smaller than those obtained with a finite electron-phonon coupling, reducing $\epsilon$ increases $T_C$, which can be justified as follows. Without electron-phonon coupling, the local attraction between electrons displayed by the Fig. \ref{Vscr} and induced by the particle-hole excitations, acts, within the present framework, as the only mechanism for the onset of superconductivity. The strength of the local attraction is sensitive to the external screening, and increases by decreasing $\epsilon$, which consequently enhances $T_C$.
\begin{figure}[h!]
\centering
\includegraphics[width=2.7in]{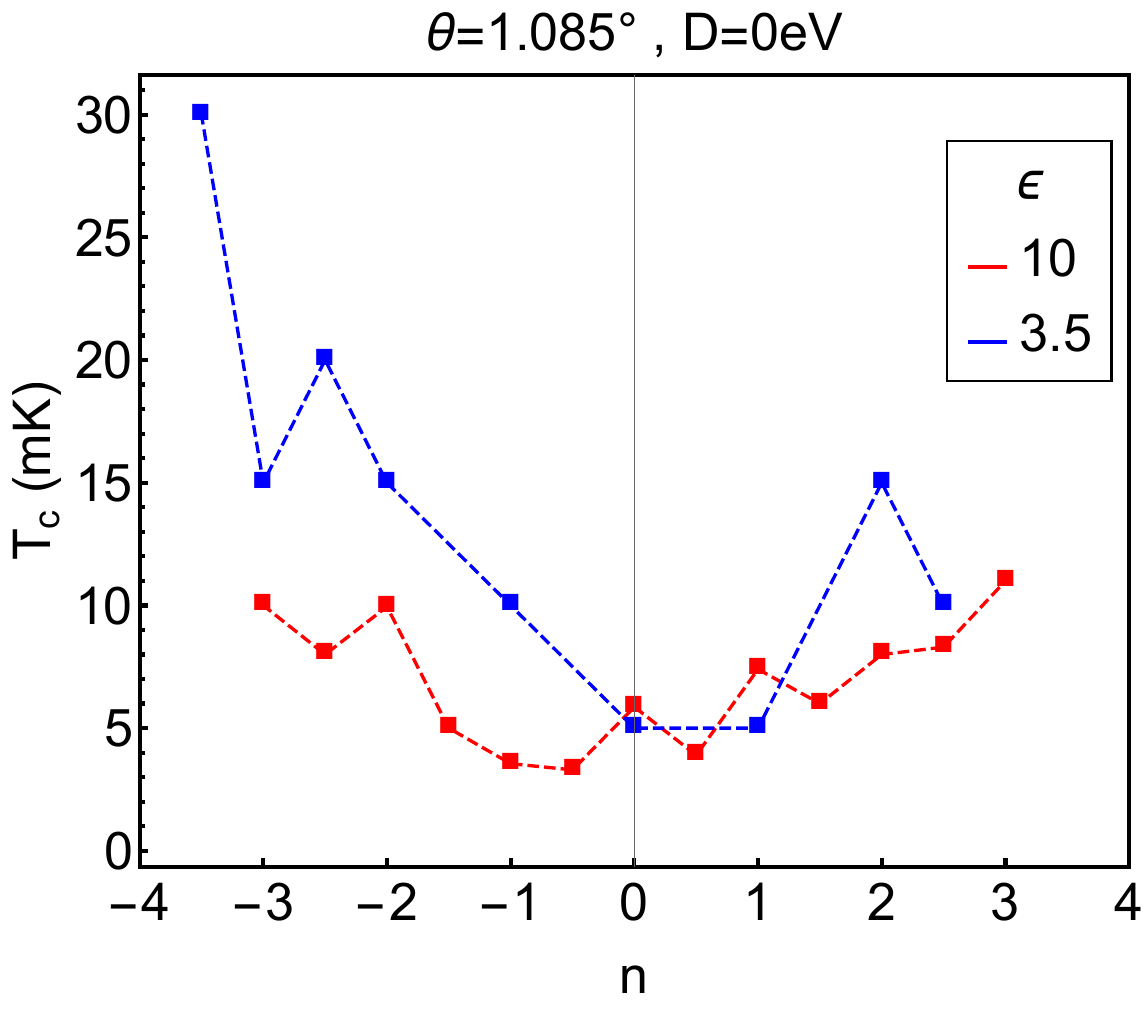}
\caption{
Critical temperatures as a function of the filling, obtained for: $\epsilon=10$ (red line) and $\epsilon=3.5$ (blue line), in the absence of electron-phonon coupling and for $\theta=1.085^\circ$.
}
\label{screening_dependence_D=0}
\end{figure}

\section{Pairing symmetry}\label{App_OP_symmetry}
The Fig. \ref{OP_symmetry} shows the amplitude of the order parameter, $\Delta$, as a function of the wave vector in the moir\'e BZ, upon varying the twist angle and the filling. The black lines identify the Fermi surfaces.
The order parameter is always concentrated in the neighboring of the Fermi surface, as expected. The only exception is the case corresponding to the magic angle, $\theta=1.085^\circ$ (a), at half filling, where the narrowness of the bands allows the order parameter to be robust in almost the whole BZ. 
Note that, at half filling and for $\theta=1.12^\circ$ (b) and $\theta=1.18^\circ$ (c), the Fermi surface collapses into the corners of the BZ, as the dispersion of the two central bands is Dirac like. 
\begin{figure*}
\centering
\includegraphics[width=6.in]{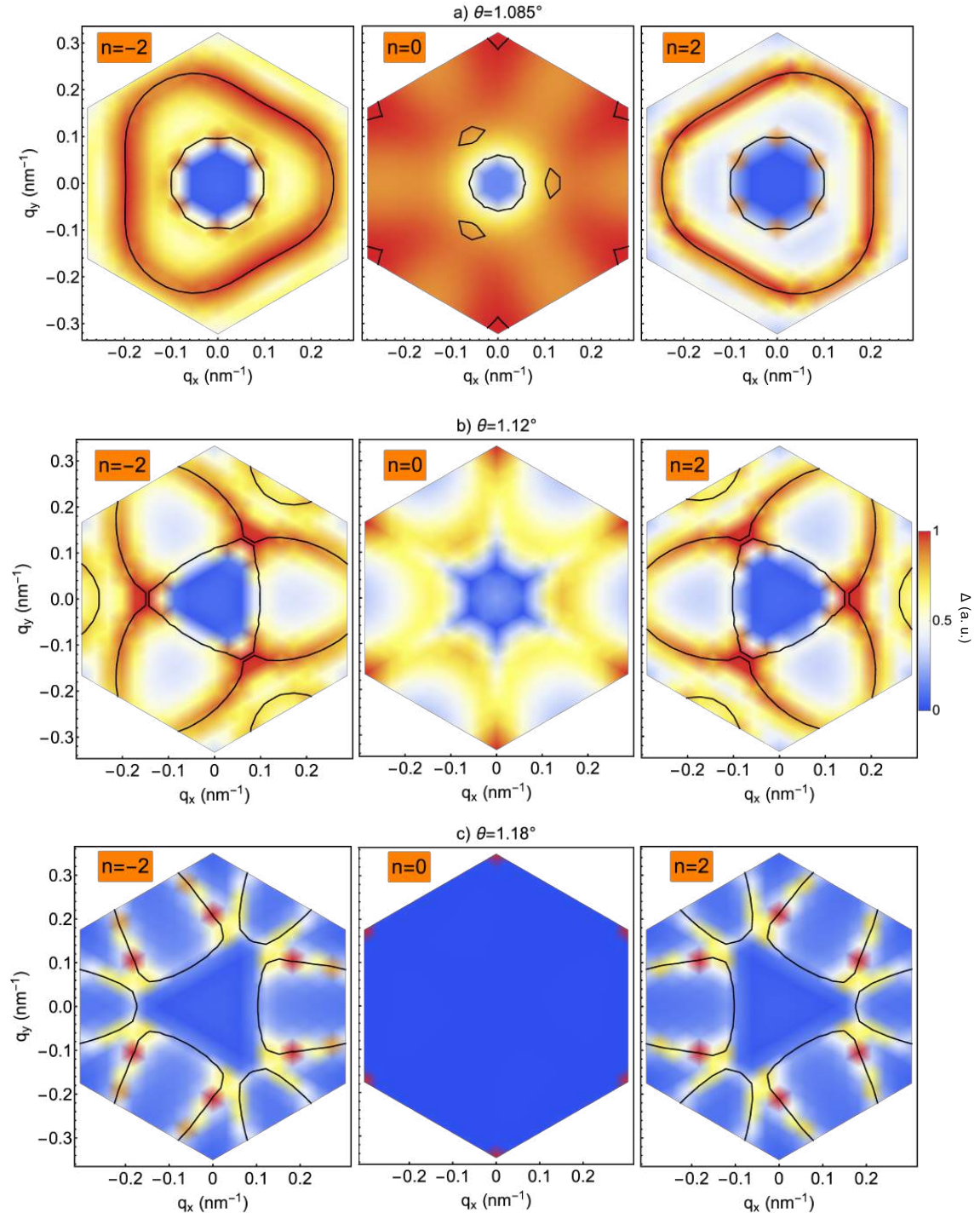}
\caption{
Amplitude of the order parameter as a function of the wave vector in the moir\'e BZ,
obtained for different angles and fillings, as denoted in the captions.
The black lines identify the Fermi surfaces.
}
\label{OP_symmetry}
\end{figure*}

\section{Analysis of the pairing potential}\label{APP_pairing_potential}
Here we study the kernel $\Gamma$, Eq. \pref{kernel}, which defines the pairing potential in the gap equation \pref{BCS_vertex_qspace}.
The Fig. \ref{fig_pairing_potential} shows the probability distribution of the real part of the matrix elements: $\Gamma^{m_1m_2}_{n_1n_2;\alpha\beta}(\vec{k},\vec{q})$, obtained for $\theta=1.085^\circ$, $n=-2$ and in three different cases:
a) in the presence of phonons, b) in the absence of phonons and c) in the absence of phonons and upon simplifying the Umklapp terms similarly to Eq. \pref{factor} (${\cal M}_{\vec{G}} ( \vec{k} , \vec{k} + \vec{q} ) = e^{- \xi^2 | \vec{G} + \vec{q} |^2}$).
We checked numerically that the imaginary parts are negligible as compared to the real parts. Remarkably, while the pairing potential can have different signs in the cases a) and b), meaning  both attractive and repulsive behavior in the reciprocal space, in the case c) the potential is always negative, meaning a completely repulsive behavior, in full agreement with the analytical estimate of the Eq. \pref{kernel2}. By comparing the Figs. \ref{fig_pairing_potential}-b) and \ref{fig_pairing_potential}-c), that do not account for the contribution of the phonons, we can then conclude that the complexity of the wave functions, encoded in the Umklapp terms, induces attraction in the reciprocal space, which favors superconductivity.
\begin{figure}
\centering
\includegraphics[width=3.in]{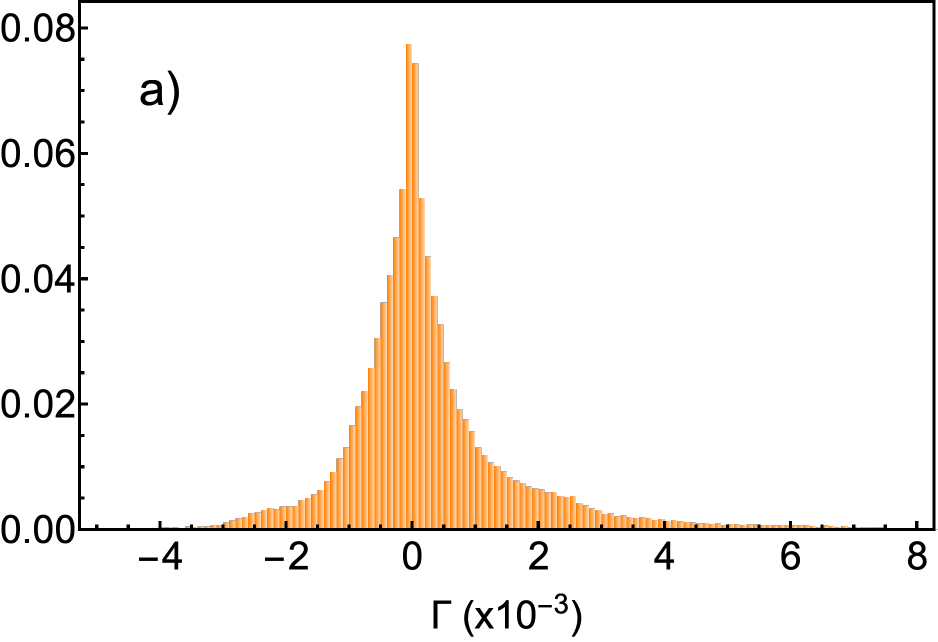}\\
\includegraphics[width=3.in]{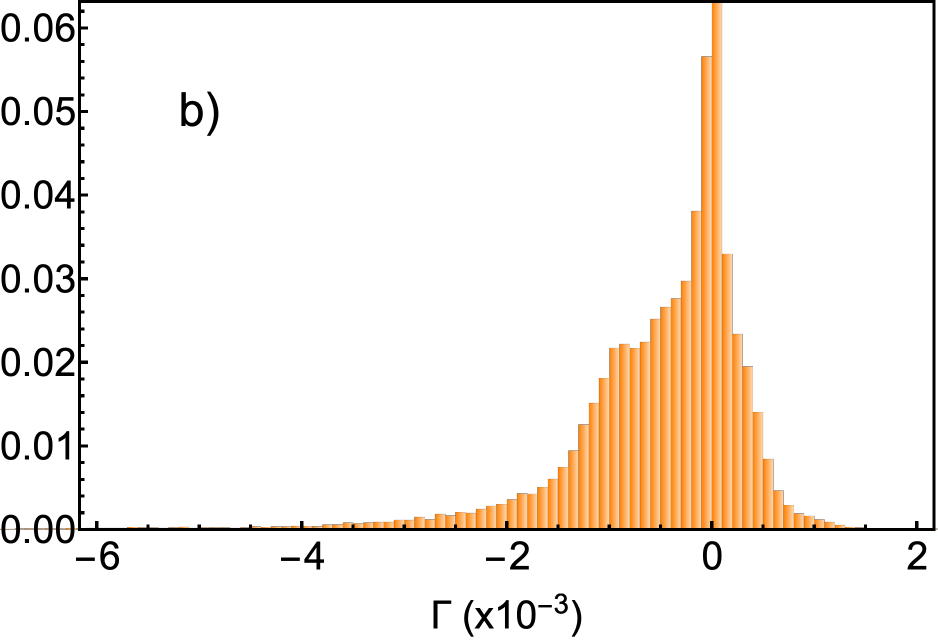}\\
\includegraphics[width=3.in]{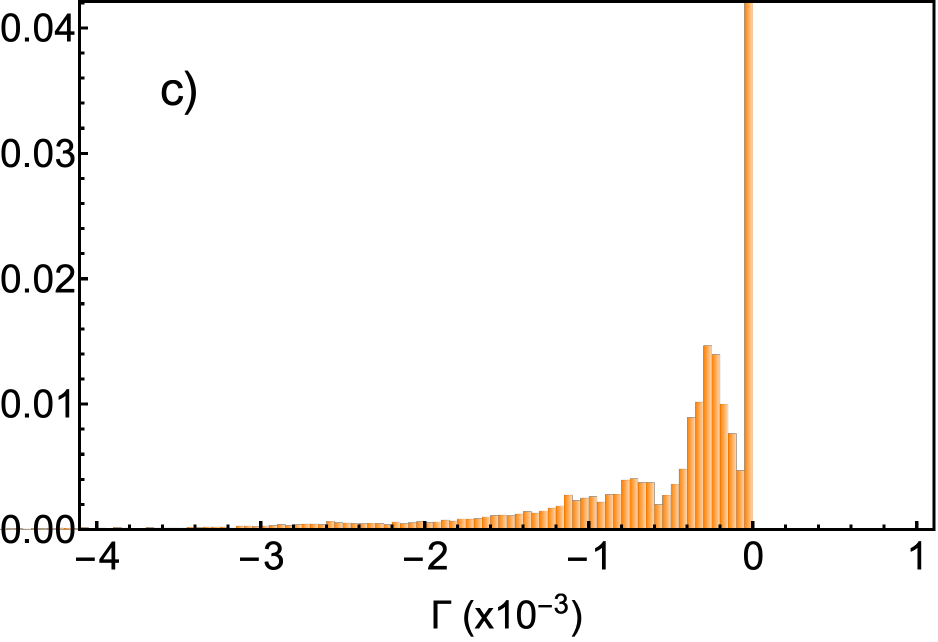}
\caption{
Probability distribution of the real part of the matrix elements: $\Gamma^{m_1m_2}_{n_1n_2;\alpha\beta}(\vec{k},\vec{q})$, obtained for $\theta=1.085^\circ$, $n=-2$ and in three different cases:
a) in the presence of phonons, b) in the absence of phonons and c) in the absence of phonons and upon simplifying the Umklapp terms according to the Eq. \pref{factor}.
}
\label{fig_pairing_potential}
\end{figure}

In the Fig. \ref{fig_symplified_Umklapp_OP_symmetry} we show the amplitude (a) and phase (b) of the order parameter corresponding to the case of simplified Umklapp terms of the Fig. \ref{fig_pairing_potential}-c). As is evident, the order parameter changes sign across the Fermi surface, a necessary consequence of the purely repulsive behavior of the pairing potential in the reciprocal space. The resulting symmetry features an angular momentum: $l>0$, in stark contrast with the robust $s$-wave gap of TBG reported so far.
\begin{figure}
\centering
\includegraphics[width=3.5in]{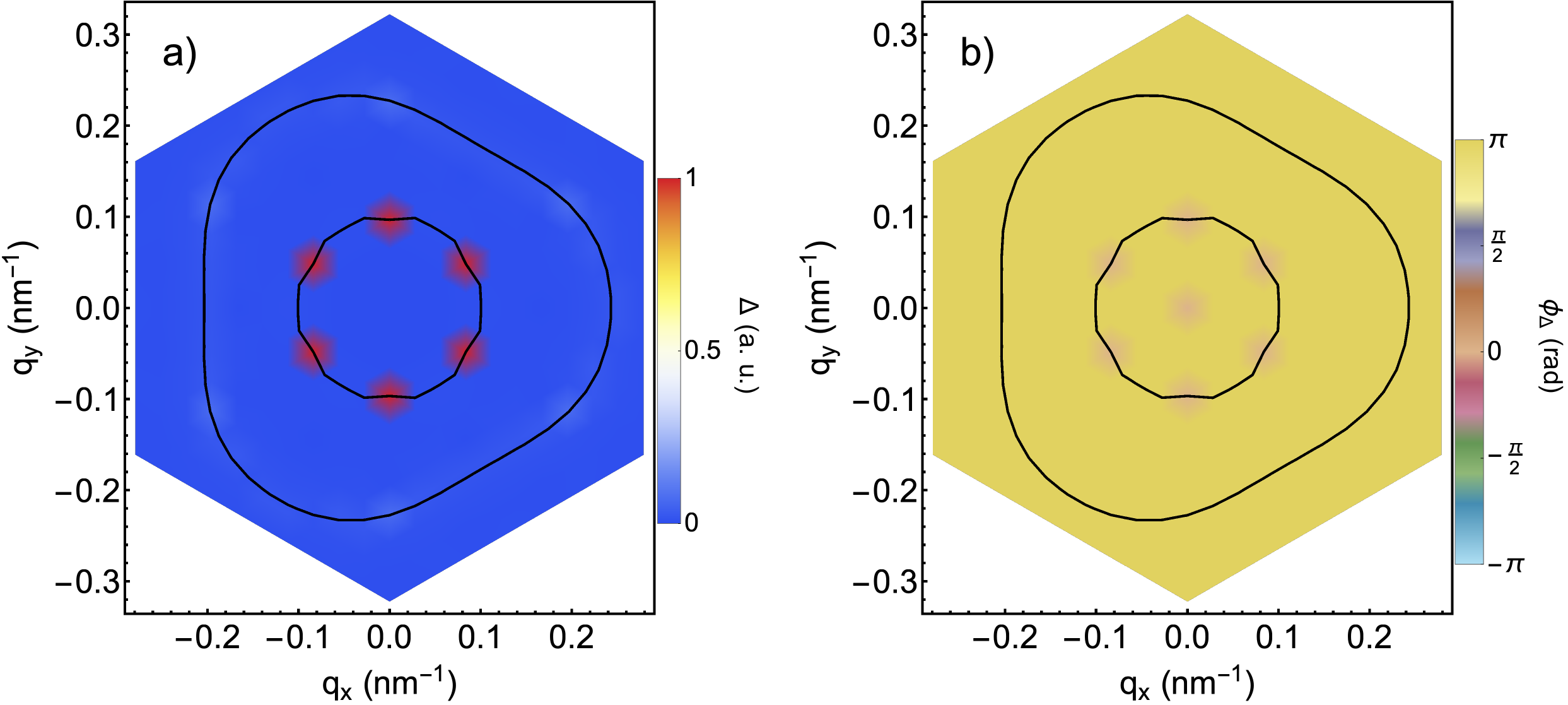}
\caption{
Amplitude (a) and phase (b) of the order parameter corresponding to the case of simplified Umklapp terms of the Fig. \ref{fig_pairing_potential}-c).
}
\label{fig_symplified_Umklapp_OP_symmetry}
\end{figure}

\bibliography{Literature}

\begin{thebibliography}{83}
\bibitem [{met()}]{metric}%
  \BibitemOpen
  \href@noop {} {}\bibinfo {note} {The quantum metric associated to the
  wavefunctions in the Brillouin Zone also affects the rigidity of the
  superconducting phase, and, as a result, it modifies the critical
  temperature, see\cite{HHPR19,JPLHT20,XSZLB20}.}\BibitemShut {Stop}%
\end{thebibliography}

\end{document}